\documentclass[twocolumn,showkeys,superscriptaddress,amsmath,amssymb,aps]{revtex4}

\usepackage{graphicx,color}
\usepackage{dcolumn}
\usepackage{bm}

\def\sech{{\rm sech}}

\begin{document}

\title{Transverse instability and dynamics of nonlocal bright solitons}

\author{G. N. Koutsokostas}
\affiliation{Department of Physics, National and Kapodistrian University of Athens, Panepistimiopolis,
Zografos, Athens 15784, Greece}

\author{G. Theocharis}
\affiliation{LAUM, CNRS, Le Mans Universit{\'e}, Avenue Olivier Messiaen,
72085 Le Mans, France}

\author{T. P. Horikis}
\affiliation{Department of Mathematics, University of Ioannina, Ioannina 45110, Greece}

\author{P. G. Kevrekidis}
\affiliation{Department of Mathematics and Statistics, University of Massachusetts,
Amherst, MA 01003-4515, USA}

\author{D. J. Frantzeskakis}
\affiliation{Department of Physics, National and Kapodistrian University of Athens,
Panepistimiopolis, Zografos, Athens 15784, Greece}

\begin{abstract}
We study the transverse instability and dynamics of bright soliton stripes in
two-dimensional nonlocal
nonlinear media. Using a multiscale perturbation method, we derive analytically the first-order
correction to the soliton shape, which features an exponential growth in time -- a signature
of the transverse instability. The soliton's  characteristic
timescale associated with its exponential growth,
is found to depend on the square
root of the nonlocality parameter. This, in turn,  highlights the nonlocality-induced suppression
of the transverse instability. Our analytical predictions are corroborated by direct numerical
simulations, with the analytical results being in good agreement with the numerical ones.

\end{abstract}

\maketitle

\section{Introduction}

Wave instabilities play an important role in the evolution of nonlinear systems,
as they are associated with symmetry breaking effects resulting in the formation of
coherent structures or leading to chaotic states \cite{Infeld}. A pertinent example is the
modulational instability (MI) of plane waves in media governed by the focusing nonlinear
Schr{\"o}dinger (NLS) equation, with MI resulting in a variety of important
nonlinear processes, such as the formation of envelope bright solitons, envelope shock waves,
and rogue waves \cite{zo}. Another example is the transverse instability, which
refers to the growth of transverse modulations of quasi one-dimensional (1D) (stripe) bright and
dark solitons, for focusing \cite{ZR,han} and defocusing \cite{kuz,peli1} NLS models,
respectively. In the elliptic 2D focusing NLS, the norm of the soliton
stripes  is infinite, as they extend to infinity in the transverse direction.
Hence, the bright soliton stripes, being subject to the onset of collapse, break up, through a
width modulation, into individual 2D lump-shaped
structures. In such a case, the transverse instability is of the so-called ``necking'' type
\cite{krz}. On the other hand, bright solitons of the hyperbolic 2D focusing NLS, as well as
dark solitons of the elliptic 2D defocusing NLS, undergo undulations
of the location of their center, due to the transverse
instability, and  eventually decay; due to this, so-called ``snaking'' instability, bright
soliton stripes decay into bright lumps, while dark solitons decay
into vortices or dark lumps \cite{pelirev,yang}. It is important to
note that this is a popular experimental technique for observing
the instability outcome and subsequent pattern formation both
in atomic physics~\cite{anderson} and in nonlinear optics~\cite{bernard,bernard2}.

Arrest or substantial suppression of the transverse instability of solitons has been proved
to be a topic of great interest, and various physical mechanisms have been proposed to suppress
this instability. These mechanisms include the coupling of solitons with another soliton
component \cite{muss1,muss2}, making the soliton sufficiently incoherent along the transverse
direction \cite{ana}, as well as using periodic lattice potentials \cite{y1,y2}
or localized barrier potentials \cite{Ma}. In addition, nonlocal nonlinearities, occurring
e.g., in plasmas \cite{litvak}, atomic vapors \cite{vap}, lead glasses \cite{rot},
nematic liquid crystals \cite{ass0}, as well as dipolar Bose-Einstein condensates \cite{santos},
are known to play a key role on the stability of solitons. In particular, in settings
with focusing nonlocal nonlinearities, the transverse instability of bright nonlocal solitons
can be substantially suppressed \cite{lin08}, while collapse can be arrested
in higher-dimensions \cite{tursolo,krol1} and, as a result, stable 2D and 3D solitons can be
formed \cite{vap,rot,Mih2,krol1,Mih}. On the other hand, in settings with defocusing nonlocal
nonlinearities, the transverse instability of dark nonlocal solitons \cite{dr1,kart1,piccardi,tph}
can be suppressed \cite{trillo}.

In this work, we revisit the problem of the transverse instability and dynamics
of bright solitons in nonlocal nonlinear media. The considered nonlocal NLS model, namely a
Schr{\"o}dinger type paraxial wave equation, coupled with a diffusion-type
equation governing the nonlocal response of the medium, is relevant to a variety
of physical contexts. These include optical media with a thermal nonlinearity (e.g.,
atomic vapors \cite{vap,rot} and liquid solutions \cite{liq1,liq2}), plasmas \cite{pl1,pl2},
and nematic liquid crystals \cite{pec,alb}. The considered nonlocal NLS possesses
a sech$^2$-shaped exact analytical bright soliton solution, which can not be reduced
-- in the local nonlinearity limit -- to the usual sech-shaped bright soliton of the NLS;
hence, one can not exploit this limit to study the effect of nonlocality on the transverse
instability of nonlocal soliton stripes, as has been done, e.g., in~\cite{lin08}.

Here, we analyze the problem by employing a
perturbation method, similar to the one used in Ref.~\cite{pes} (see also \cite{MJA.segur}),
which uses a perturbation ansatz relying on the sech$^2$ nonlocal soliton, with a center
and a phase becoming unknown functions of slow time and transverse coordinate. We find an
approximate solution of the nonlocal NLS, where the correction to the soliton
shape is shown to feature an exponential growth in time, which is a signature of
the transverse instability. The latter is of the necking type, and is induced by the soliton
phase, which is shown to obey an elliptic partial differential equation (PDE).
We show that the instability growth rate (i.e., the inverse characteristic time scale
associated with the exponential growth of the phase), depends on the inverse square root
of the nonlocality parameter, a fact highlighting the substantial, nonlocality-induced
suppression of the transverse instability of the bright soliton stripes
(in a similar vein as earlier works \cite{lin08}).
The analytical estimation for the growth rate, as well as the derived approximate
analytical solution, are found to be in good
agreement with respective results obtained by means of direct simulations.
Despite the prolongation of the lifetime of the solutions obtained
herein, our results do not support the scenario of a complete stabilization
of the relevant bright soliton stripes, irrespectively of the value of
nonlocality parameter $\nu$, for the range considered herein.

The presentation of the manuscript is organized as follows. In Section~II, we introduce the model
and its exact soliton solution, and present the results of our perturbation method; these include
the derivation of the evolution of the soliton parameters, the derivation of the
instability growth rate, as well as the first-order correction
of the soliton shape. Section~III is devoted to the presentation of our numerical
results, and comparison with the analytical approximations. Finally, in Section~IV we summarize
our conclusions and discuss possibilities for relevant future research.

\section{Model and stability analysis}
\subsection{The model and its exact $1$D soliton solutions}

We consider the propagation of an optical beam in a nonlocal nonlinear medium. Let $u$
be the complex electric field envelope of the light beam satisfying a paraxial,
Schr{\"o}dinger-type equation, and the
real function $\theta$ be the nonlinear, generally nonlocal, medium's
response, assumed to obey a diffusion-type equation~\cite{Mih}. Then, the evolution of the
beam is governed by the following dimensionless nonlocal NLS model:
\begin{eqnarray}
i u_t + \frac{d}{2} \Delta u + 2 g \theta u=0,
\label{1} \\
\nu \Delta \theta -2q \theta + 2g |u|^2=0,
\label{2}
\end{eqnarray}
where subscripts denote partial derivatives. Here, the evolution variable $t$ represents
the propagation distance (assumed to be along the $z$ direction),
$\Delta\equiv \partial_x^2+\partial_y^2$ is the
transverse Laplacian, while $g$ and $d$ are coupling and diffraction coefficients,
assumed to be positive; this case corresponds to a focusing nonlinearity. In addition,
$q>0$ is a constant and, finally, the parameter $\nu$, which measures the diffusion length
(assumed to be large compared to the operating wavelength), describes the strength of
nonlocality: indeed, large $\nu$ corresponds to a highly nonlocal response while in the limit
$\nu \rightarrow 0$, Eqs.~(\ref{1})-(\ref{2}) reduce to the following NLS equation
with a local cubic (Kerr-type) nonlinearity:
\begin{eqnarray}
i u_t + \frac{d}{2} \Delta u + \frac{2g^2}{q} |u|^2 u=0.
\label{3}
\end{eqnarray}
The model~(\ref{1})-(\ref{2}) is relevant to a variety of nonlocal media.
These include: (a) optical media featuring a thermal nonlinearity -- such as atomic
vapors \cite{vap,rot} and liquid solutions, with $\theta$ being the nonlinear
correction to the refractive index \cite{liq1,liq2}; (b) ionized plasmas,
with $\theta$ being the relative electron
temperature perturbation, and $q \propto m/M$ being the relative energy
that an electron of mass $m$ delivers to a heavy particle of mass $M$ during
a single collision \cite{pl1,pl2}; (c) nematic liquid
crystals \cite{pec,alb}, with $\theta$ denoting the perturbation of the optical
director angle from its static value due to the light beam, and
$q$ being related to 
the applied static field which pretilts the nematic
dielectric \cite{ass0,ass1}.

As explained in the Introduction, our scope is to study the transverse dynamics of
1D bright soliton stripes in the 2D setting and investigate, in particular, the role
of nonlocality. It is thus convenient to start by presenting such 1D bright soliton
solutions of Eqs.~(\ref{1}-\ref{2}), which can be found upon using the ansatz \cite{pana2}:
\begin{equation}
u=q_0(\xi)\exp[i\omega(t+\sigma_0)], \quad
\theta = \theta_0(\xi),
\label{ans}
\end{equation}
where $q_0$ is an unknown real function depending on $\xi = k (x-x_0)$,
$k$ is an unknown constant, $\omega$ is the unknown frequency of the solution, while
$x_0$ and $\sigma_0$ are arbitrary real parameters representing, respectively,
the initial location and the phase of the soliton. Substituting Eqs.~(\ref{ans}) into
Eqs.~(\ref{1})-(\ref{2}), it can be found that the resulting equations become:
\begin{eqnarray}
d k^2 q_{0\xi\xi}-2\omega q_0+4g q_0\theta_{0}&=&0,
\label{ode1}\\
\nu k^2 \theta_{0\xi\xi}-2q\theta_{0}+2g q_0^2&=&0.
\label{ode2}
\end{eqnarray}
Then, observing that if
\begin{equation}
\theta_0=\sqrt{\frac{d}{2\nu}}q_0, \quad {\rm and} \quad \omega =\frac{d q}{\nu},
\label{thq}
\end{equation}
then the system~(\ref{ode1})-(\ref{ode2}) reduces to a single ordinary
differential equation (ODE):
\begin{equation}
q_{0\xi\xi} - \frac{2g}{\nu k^2}q_0 + \frac{4g}{\sqrt{2\nu d}k^2} q_0^2 =0.
\end{equation}
The latter possesses the exact soliton solution
\begin{equation}
q_0(\xi)= \frac{3q}{2g}\sqrt{\frac{d}{2\nu }} \sech^{2}\left[k (x-x_0)\right],
\quad
k=\sqrt{\frac{q}{2\nu}},
\label{q0}
\end{equation}
which implies that the soliton solutions of Eqs.~(\ref{1})-(\ref{2})
are of form:
\begin{eqnarray}
\!\!\!\!\!\!\!\!\!
u_0(x,t)&=&\frac{3q}{2g}\sqrt{\frac{d}{2\nu }}
\sech^{2}\left[k (x-x_0)\right]\exp[i\omega (t + \sigma_0)],
\label{4}\\
\!\!\!\!\!\!\!\!\!
\theta_0(x,t)&=&\frac{3 d q}{4 g \nu} \sech^{2}\left[k (x-x_0)\right].
\label{5}
\end{eqnarray}
Here it is interesting to note that, while the system of Eqs.~(\ref{1})-(\ref{2})
reduces to the local NLS~(\ref{3}), the exact solution~(\ref{4}) cannot be reduced
to the soliton solution of Eq.~(\ref{3}) (which features a sech-profile and is
characterized by a free parameter); this becomes clear by the fact that
$\lim_{\nu \rightarrow 0}u = 0$. It is also mentioned that
Eqs.~(\ref{4})-(\ref{5}) represent a stationary solution of the problem; traveling
solutions exist as well and can easily be constructed by means of a Galilean boost.

\subsection{Perturbation theory}

In order to study the stability of solutions (\ref{4})-(\ref{5}) in two dimensions, we
consider solutions of Eqs.~(\ref{1})-(\ref{2}) in the form of the following asymptotic
expansions:
\begin{eqnarray}
\!\!\!\!\!\!\!\!\!
u(\xi,t,T_{i},Y_{i})&=& \sum_{j=0}^{\infty} \epsilon^{j} 
q_j(\xi)\exp\left[i\omega\left(t+\sigma_{0}(T_{i},Y_{i})\right)\right],
\label{6}
\\
\!\!\!\!\!\!\!\!\!
\theta(\xi,T_{i},Y_{i})&=&\sum_{j=0}^{\infty} \epsilon^{j} \theta_j(\xi,T_{i},Y_{i}),
\label{7} \\
\xi &=&k\left[x-x_{0}(T_{i},Y_{i})\right]
\end{eqnarray}
where $T_{i}=\epsilon^{i}t$, $Y_{i}=\epsilon^{i}y$ and $0<\epsilon\ll1$ is a formal
small parameter. The above perturbation ansatz is actually inspired by the form of the
exact solution~(\ref{ans}) of the 1D problem but, now, with the soliton's center $x_0$ and
phase $\sigma_0$ becoming unknown functions of the slow variables $T_{i}$ and $Y_{i}$.
Substituting the perturbation expansions (\ref{6})-(\ref{7}) into Eqs. (\ref{1})-(\ref{2})
we obtain the following results.

First, at $\mathcal{O}(\epsilon^{0})$, we obtain the system~(\ref{ode1})-(\ref{ode2}), which
provides the exact soliton solution (\ref{q0}) [that eventually leads, together with (\ref{thq}),
to the solution of Eqs.~(\ref{4})-(\ref{5})].

At the next orders of approximation, the presence of derivatives of $x_0$ and $\sigma_0$ with
respect to the slow variables renders the inhomogeneous parts $F_j$ of the resulting
equations for $q_j$ (with $j=1,2,\ldots$) complex, i.e., $F_j=F_j^{(\rm r)}+iF_j^{(\rm i)}$;
this implies that $q_j$ itself must be complex, i.e., $q_j=q_j^{(\rm r)}+iq_j^{(\rm i)}$.
Thus, separating real and imaginary parts of the resulting equations, we obtain,
at each order, a set of three equations, two of which are coupled.
To be more specific, the resulting equations at orders $\mathcal{O}(\epsilon^{j})$ for
$j=1,2,\ldots$ take the following form:
\begin{eqnarray}
\!\!\!\!\!\!\!
&&\left(d k^2 \partial_{\xi}^2-2\omega +4g\theta_{0}\right)q_{j}^{(\rm r)}
+4g q_0\theta_{j}= F_j^{(\rm r)},
\label{14} \\
\!\!\!\!\!\!\!
&&\left(d k^2 \partial_{\xi}^2 -2\omega +4g\theta_{0}\right)q_{j}^{(\rm i)} =
F_j^{(\rm i)}
\label{15}\\
\!\!\!\!\!\!\!
&&\left(\nu k^2 \partial_{\xi}^2 -2q\right)\theta_{j}+4g q_0q_{j}^{(\rm r)}=G_j.
\label{16}
\end{eqnarray}
The inhomogeneous parts at the order $\mathcal{O}(\epsilon)$ are given by:
\begin{equation}
F_1^{(\rm r)}= 2\omega \sigma_{0T_1} q_0, \quad
F_1^{(\rm i)}= 2 k  x_{0T_1} q_{0\xi}, \quad
G_1 = 0,
\end{equation}
while at the order $\mathcal{O}(\epsilon^2)$ they read:
\begin{eqnarray}
F_2^{(\rm r)}=&-&4gq_1^{(\rm r)}\theta_{1}-d k^2 x_{0Y_1}^{2}q_{0\xi\xi}
+ d k x_{0Y_1Y_1}q_{0\xi}
\nonumber\\
&+&d\omega^{2}\sigma_{0Y_1}^{2}q_0 + 2\omega\sigma_{0T_2}q_{0}+2\omega\sigma_{0T_1}q_1^{(\rm r)}
\nonumber\\
&+&2q_{1T_1}^{(\rm i)}-2vx_{0T_1}q_1^{(\rm i)},
\label{23}
\\
F_2^{(\rm i)}=&-&4g q_1^{(\rm i)}\theta_{1}+2d k \omega x_{0Y_1}\sigma_{0Y_1}q_{0\xi}
-d\omega\sigma_{0Y_1Y_1}q_{0}
\nonumber\\
&+&2kx_{0T_2}q_{0\xi}+2\omega\sigma_{0T_1}q_1^{(\rm i)}-2q_{1T_1}^{(\rm r)}
\nonumber \\
&+&2k x_{0T_1} q_{1\xi}^{(\rm r)},
\label{24}
\\
G_{2}=&-&\nu k^2 x_{0Y_1}^{2}\theta_{0\xi\xi}+\nu k x_{0Y_1Y_1} \theta_{0\xi}
\nonumber \\
&-&2g\left(q_{1}^{(\rm r)2}+q_{1}^{(\rm i)2}\right).
\label{25}
\end{eqnarray}

To proceed further, it is useful to make a few observations. First, differentiating Eqs.~(\ref{ode1})-(\ref{ode2}) with respect to $\xi$, one obtains the homogeneous part
of Eqs.~(\ref{14})-(\ref{16}). This implies that the homogeneous
solutions of Eqs.~(\ref{14})-(\ref{16}) are of the form:
\begin{eqnarray}
q_{jh}^{(\rm r)}=q_{0\xi}, \quad q_{jh}^{(\rm i)}=q_0, \quad \theta_{jh}=\theta_{0\xi}.
\label{26}
\end{eqnarray}
Second, having found the above homogeneous solutions, we may derive
the solvability conditions of the full inhomogeneous problem, Eqs.~(\ref{14})-(\ref{16}).
To do this, first we consider the coupled Eqs.~(\ref{14}) and (\ref{16}).
We multiply both sides of Eqs.~(\ref{14}) by the homogeneous solution
$q_{jh}^{(\rm r)}$, as well as both sides of Eq.~(\ref{16}) by
the homogeneous solution $\theta_{jh}$. Then, we add the resulting equations and
integrate with respect to $\xi$ from $-\infty$ to $+\infty$. This yields the following
integral relation:
\begin{eqnarray}
\int_{-\infty}^{\infty}\left(q_{jh}^{(\rm r)}F_{j}^{(\rm r)}
+\theta_{jh} G_{j}\right) d\xi=0,
\label{29}
\end{eqnarray}
which is the solvability condition of Eqs.~(\ref{14}) and (\ref{16}). To obtain the
solvability condition for Eq.~(\ref{15}), we follow a similar procedure, namely
we multiply both sides of Eq.~(\ref{15}) by the homogeneous solution
$q_{jh}^{(\rm i)}$ and integrate from $-\infty$ to $+\infty$; this yields:
\begin{eqnarray}
\int_{-\infty}^{\infty} q_{jh}^{(\rm i)}F_{j}^{(\rm i)} d\xi =0.
\label{30}
\end{eqnarray}
Importantly, the above solvability conditions will lead to evolution equations for the
soliton center $x_{0}$ and phase $\sigma_{0}$ which ---as we will see--- will provide
the necessary information for characterizing the stability of the 1D soliton solutions.
Furthermore, solving Eqs.~(\ref{14})-(\ref{16}) (for $j=1$) will provide us the form
of the solution to Eqs.~(\ref{1})-(\ref{2}) up to $\mathcal{O}(\epsilon)$, and for
short times -- up to the onset of the instability. This will be particularly relevant
for our direct numerical simulations as well.

\subsection{Evolution of the soliton parameters}

First we consider the problem at the order $\mathcal{O}(\epsilon)$. In this case, the
solvability conditions, Eqs.~(\ref{29}) and (\ref{30}), lead to the following results,
respectively:
\begin{equation}
\int_{-\infty}^{\infty}q_{0\xi}\left(2\omega \sigma_{0T_1} q_0 \right)d\xi=2\omega \sigma_{0T_1}  \int_{-\infty}^{\infty}\left(q_{0} q_{0\xi}\right)d\xi=0,
\label{31}
\end{equation}
and
\begin{equation}
\int_{-\infty}^{\infty}q_{0} \left(2v x_{0T_1} q_{0\xi}\right)d\xi=2k x_{0T_1}  \int_{-\infty}^{\infty}\left(q_{0} q_{0\xi}\right)d\xi=0.
\label{32}
\end{equation}
The above results indicate that the solvability conditions at this order, $\mathcal{O}(\epsilon)$,
are always satisfied, regardless of the specific form of the soliton parameters $x_0$ and
$\sigma_0$. Thus, we may proceed by solving Eqs.~(\ref{14})-(\ref{16})
(see details for the derivation of these solutions in the Appendix), and find the
following exact solutions for the soliton correction $q_1$:
\begin{eqnarray}
q_{1}^{(\rm i)}&=&\frac{3}{2g\sqrt{d}}x_{0T_1} \xi \sech^{2}\left(\xi\right),
\label{33}\\
q_1^{(\rm r)} &=&-\frac{3q}{16g}\sqrt{\frac{d}{2\nu}}\sigma_{0T_1}\sech^3(\xi)\Bigg[-9\cosh(\xi)
\nonumber\\
&+&\cosh(3\xi)+12\xi\sinh(\xi)\Bigg].
\label{34}
\end{eqnarray}
Notice that $q_1^{(\rm r)} \rightarrow
-\frac{3q}{16g}\sqrt{\frac{d}{2\nu}}\sigma_{0T_1}$ as $|\xi| \rightarrow \infty$, a fact that
is associated with the emergence of a {\it shelf}, i.e., a linear wave adjacent to the soliton.
Shelves were first found in the context of perturbed Korteweg-de Vries (KdV) equations
\cite{MJA.segur}, and later were also studied for both focusing \cite{kath} and defocusing
\cite{mjads} NLS models with a local nonlinearity. Generally, the emergence of shelves lead
to the breakdown of the perturbation theory at a higher order approximation in the perturbation
scheme \cite{MJA.segur}. While this issue, along with the appearance of the shelf, are
interesting by themselves, they will not be considered here; in our case, the instability
induced by the presence of $\sigma_{0T_1}$ in Eq.~(\ref{34}) (see below) plays the dominant
role in the evolution of the soliton.

To proceed further, we apply the solvability condition at $\mathcal{O}(\epsilon^{2})$, in which case Eqs.~(\ref{29}) and (\ref{30}) respectively read:
\begin{eqnarray}
\int_{-\infty}^{\infty} &\Big[&q_{0\xi}\Big(dkx_{0Y_1Y_1}q_{0\xi}
+2\omega\sigma_{0T_2}q_{0}
\nonumber \\
&+&2q_{1T_1}^{(\rm i)}\Big)+\theta_0\left(\nu k x_{0Y_1Y_1}\eta_{0\xi}\right)\Big]d\xi =0
\nonumber\\
&\Rightarrow& x_{0T_1T_1}-\frac{3d^2q}{5\nu}x_{0Y_1Y_1}=0,
\label{35}
\end{eqnarray}
and
\begin{eqnarray}
&&\int_{-\infty}^{\infty} \left[ q_{0} \left( -d\omega \sigma_{0Y_1Y_1}q_{0}
+2kx_{0T_2}q_{0\xi}-2q_{1T_1}^{(\rm r)}\right)\right]d\xi = 0
\nonumber\\
&& \qquad \qquad ~\Rightarrow  \sigma_{0T_1T_1}+\frac{4d^2q}{3\nu}\sigma_{0Y_1Y_1}=0.
\label{36}
\end{eqnarray}
The set of Eqs.~(\ref{35}) and (\ref{36}), which is one of the the main results of our
analytical approach, must be satisfied in order for our original system of
Eqs.~(\ref{14})-(\ref{16}) to be solvable, up to the order $\mathcal{O}(\epsilon^2)$.
We note that, in the above equations, no nonlinear terms in $x_0$ [in (\ref{35})]
and $\sigma_0$ [in (\ref{36})] are involved, since such terms vanish when applying
the solvability condition; thus, Eqs.~(\ref{35}) and (\ref{36}) do not
involve any approximation, up to this order of approximation.

Evidently, Eq.~(\ref{35}) is a hyperbolic PDE
(having the form of the usual 2nd-order wave equation) and, thus, its solutions corresponding
to bounded initial data never blow up. On the contrary, Eq.~(\ref{36}) is an elliptic PDE (of the
Laplace type) and, thus, any bounded initial condition features an exponential growth.
As a consequence, the exponential growth of $\sigma_0$ will result in an exponential growth
of $q_1^{(\rm r)}$ as indicated by Eq.~(\ref{34}); in other words, any initial condition
of the form~(\ref{4})-(\ref{5}) is unstable in the 2D setting.
Notice that the fact that $x_0(Y_1,T_1)$ obeys a hyperbolic PDE, while
$\sigma_0(Y_1,T_1)$ obeys an elliptic PDE, bears resemblance to the case of the instability of
bright soliton stripes of
the elliptic NLS equation, with local cubic nonlinearity, in $(2+1)$-dimensions
\cite{MJA.segur,yang}.

\subsection{Instability and instability growth rate}

To investigate the instability-induced soliton dynamics, first we note that,
in practice, the instability is anticipated to manifest itself at finite time.
This means that there exists a  characteristic timescale $\tau$ for the manifestation of the
instability leading the bright soliton stripe to decay into purely 2D structures
(similarly to the case of the elliptic 2D NLS \cite{yang}). To calculate this timescale,
we need to consider some specific initial conditions
for the PDEs~(\ref{35}) and (\ref{36}). In particular, without loss of generality, we
supplement Eq.~(\ref{35}) with the following initial data:
\begin{eqnarray}
x_{0}(0,Y_1)=\delta \cos(KY_1), \quad x_{0T_1}(0,Y_1)=0,
\label{ic1}
\end{eqnarray}
and Eq.~(\ref{36}) with the initial data:
\begin{eqnarray}
\sigma_{0}(0,Y_1)=\delta \cos(KY_1), \quad \sigma_{0T_1}(0,Y_1)=0,
\label{ic2}
\end{eqnarray}
where $\delta$ and $K$ represent the perturbation amplitude and wavenumber, respectively.
Then, the solutions of Eqs.~(\ref{35})-(\ref{36}) take, respectively, the following form:
\begin{eqnarray}
x_{0}(T_1,Y_1)&=&\frac{\delta}{2}\Bigg\{\cos \left[K\left(Y_1-\sqrt{\frac{3d^2q}{5\nu}}T_1\right)
\right]
\nonumber \\
&+&\cos \left[K\left(Y_1+\sqrt{\frac{3d^2q}{5\nu}}T_1\right)\right]\Bigg\},
\label{ics2} \\
\sigma_{0}(T_1,Y_1)&=&\frac{\delta}{2}\Bigg[ \exp\left( K\sqrt{\frac{4d^2q}{3\nu}}T_1\right)
\cos\left(K Y_1\right)
\nonumber \\
&+&\exp\left(-K\sqrt{\frac{4d^2q}{3\nu}}T_1\right)\cos\left(KY_1\right)\Bigg].
\label{ics1}
\end{eqnarray}
Obviously, Eq.~(\ref{ics2}) represents the usual D'Alembert solution composed by a
right- and a left-going wave; this solution is always bounded and never grows.
This, however, is not the case of the solution~(\ref{ics1}), which grows exponentially.
In fact, it can be inferred from Eq.~(\ref{ics1}) that the solution grows in time as:
$\sigma_{0} \propto \exp(t/\tau)$, where the characteristic time scale $\tau$ is given by:
\begin{equation}
\tau = \frac{1}{\Gamma}\equiv \frac{1}{\epsilon K \sqrt{\frac{4d^2q}{3\nu}}}
=\frac{1}{2\epsilon Kd}\sqrt{\frac{3\nu}{q}},
\label{38}
\end{equation}
with $\Gamma = 1/\tau$ being the instability growth rate.
It is important to point out that Eq.~(\ref{38}) reveals that, for fixed $\epsilon$, $K$,
$d$ and $q$, the characteristic time $\tau$ scales according to the $\sqrt{\nu}$ law, meaning
that the instability manifests itself for longer times as the nolocality becomes stronger.
As mentioned in the Introduction, suppression of instabilities is a generic feature
of nonlocality; this occurs in our case as well, but the transverse instability of the bright soliton stripe cannot be completely arrested. Nevertheless,
strong nonlocality (i.e., large $\nu$) is able to significantly prolong the soliton lifetime.

To further quantify the above results, and as a preamble for our numerical simulations, it
is now convenient to write down the soliton solution, up to $\mathcal{O}(\epsilon)$, namely:
$$u(x,y,t)= u_0(\xi)+\epsilon u_1(\xi,T_1,Y_1)+\mathcal{O}(\epsilon^2),$$
with $\xi=k[x-x_0(T_1,Y_1)]$. Substituting Eq.~(\ref{33}), Eq.~(\ref{34}) and Eq.~(\ref{4}) into  the expression above the soliton solution takes the form:
\begin{eqnarray}
u(x,y,t)&=&\Bigg\{\frac{3q}{2g}\sqrt{\frac{d}{2\nu }}
\sech^{2}(\xi) +\epsilon\Bigg[-\frac{3q}{16g}\sqrt{\frac{d}{2\nu}}
\nonumber \\
&\times &\sigma_{0T_1} \sech^3(\xi)\Big(-9\cosh(\xi)+\cosh(3\xi)
\nonumber\\
&+&12\xi\sinh(\xi)\Big)+i\frac{3}{2g\sqrt{d}}x_{0T_1} \xi \sech^{2}\left(\xi\right)\Bigg]\Bigg\}
\nonumber\\
&\times & \exp[i\omega (t + \sigma_0)] + \mathcal{O}(\epsilon^2),
\label{sol}
\end{eqnarray}
with $\omega$ given in Eq.~(\ref{thq}).
It is now clear that the soliton solution $u(x,y,t)$ grows exponentially due to the
presence of the term $\sigma_{0T_1}$ and eventually will break up. In the next Section, we will present numerical results to study the
instability dynamics, check the validity of the  solution~(\ref{sol}), as well as the estimation for the growth rate [Eq.~\ref{38})] against direct numerical simulations.

\section{Numerical results}

We now proceed with results obtained by means of dynamical simulations of
the system's evolution. The latter are
performed by numerically integrating Eqs.~(\ref{1})-(\ref{2}) using a high accuracy spectral method \cite{kassam}. The initial condition is borrowed from the soliton solution of Eq.~(\ref{sol}), for $t=0$, using the initial conditions of Eqs.~(\ref{ic1})-(\ref{ic2}). In particular, the initial condition for the field $u$
is taken to be:
\begin{eqnarray}
u(x,y,0)&=& \Bigg\{ \frac{3q}{2g}\sqrt{\frac{d}{2 \nu}}
\sech^{2}\Big[k\Big(x-\delta \cos(\epsilon K y)\Big)\Big]\Bigg\}
\nonumber\\
&\times & \exp[i\omega \delta \cos(\epsilon K y)].
\label{39}
\end{eqnarray}
It is clear that the terms $\propto \delta$ above describe  small
perturbations in the initial soliton center position and phase, while
the argument of the soliton's phase, which is $\propto \epsilon$, implies
that the relevant perturbation is a long-wavelength one.

The parameter values used in the simulations are:
\begin{eqnarray}
q=d=g=1, \quad K=3, \quad \epsilon=\delta=0.1
\label{para}
\end{eqnarray}
while the nonlocality parameter $\nu$ was varied in the interval $[1,20]$.
Notice that both $\epsilon$ and $\delta$, which were considered as small parameters
in our perturbation scheme, were assumed to take relatively large values; nevertheless,
as we will see, even for such a choice, the analytical results are found to be in
good agreement with the results of the simulations. It is also noticed that other choices
for the rest of the parameter values led to results qualitatively similar to the ones
that will be presented below.


\begin{figure}[tbp!]
\centering
\includegraphics[height=3.41cm]{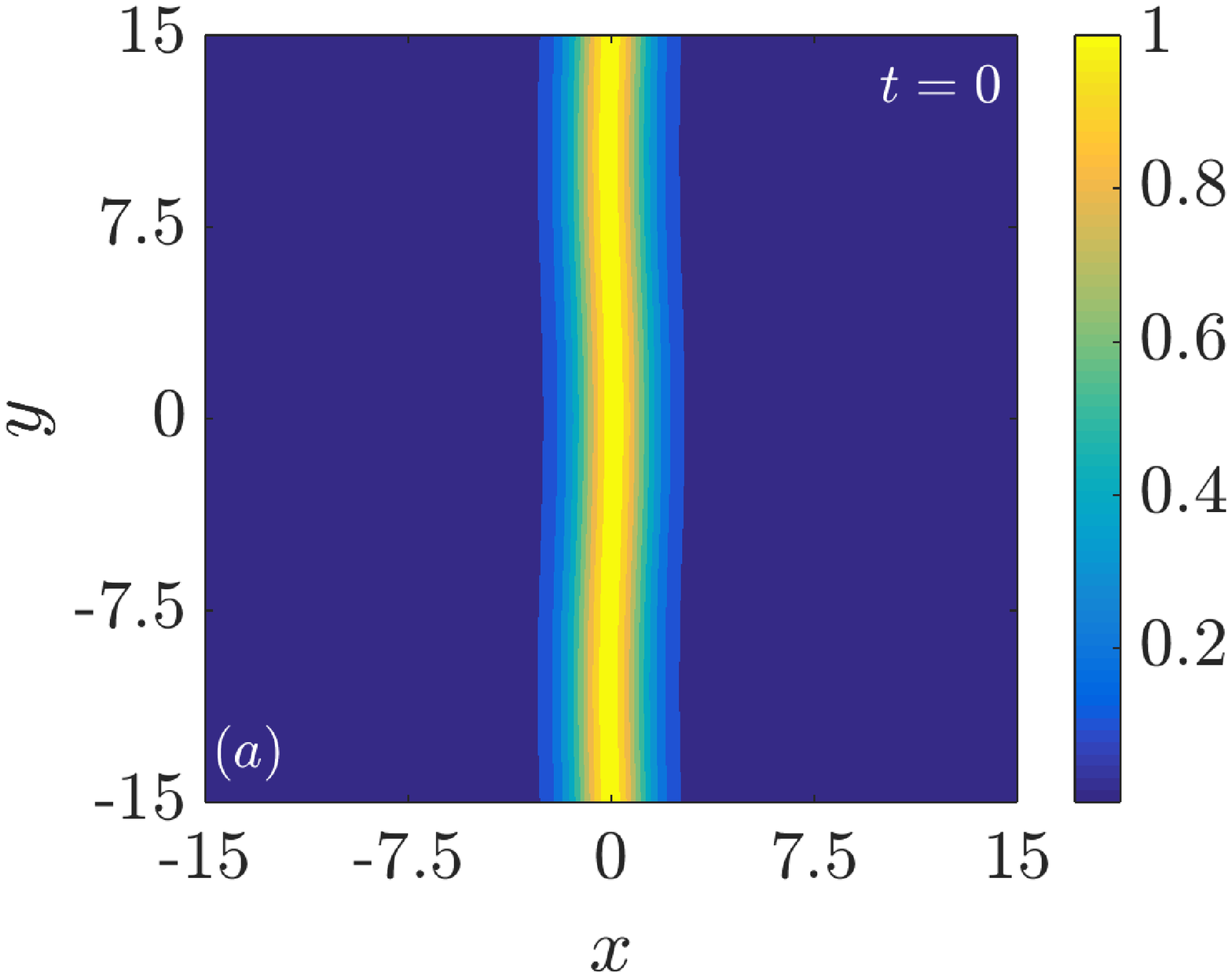}
\includegraphics[height=3.41cm]{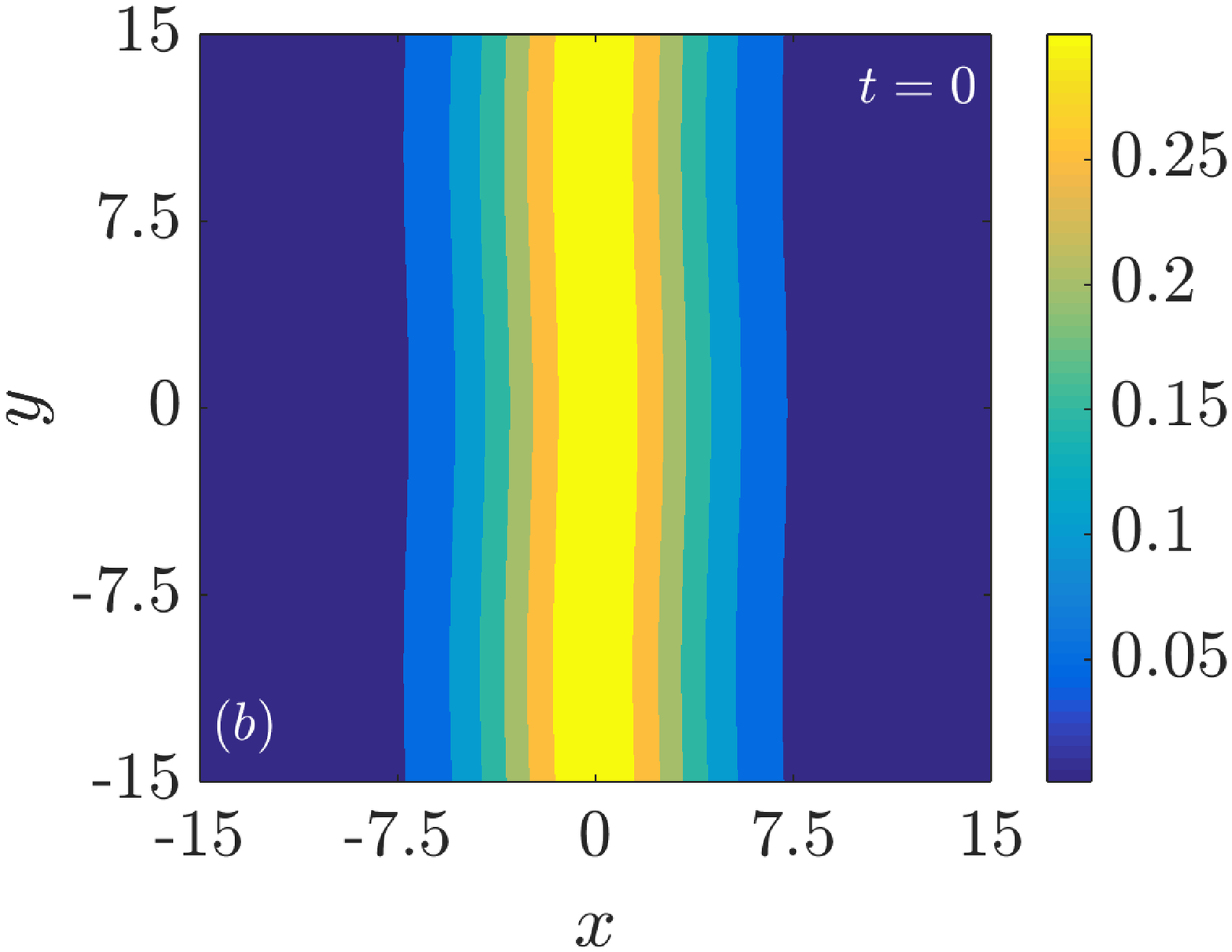}
\includegraphics[height=3.41cm]{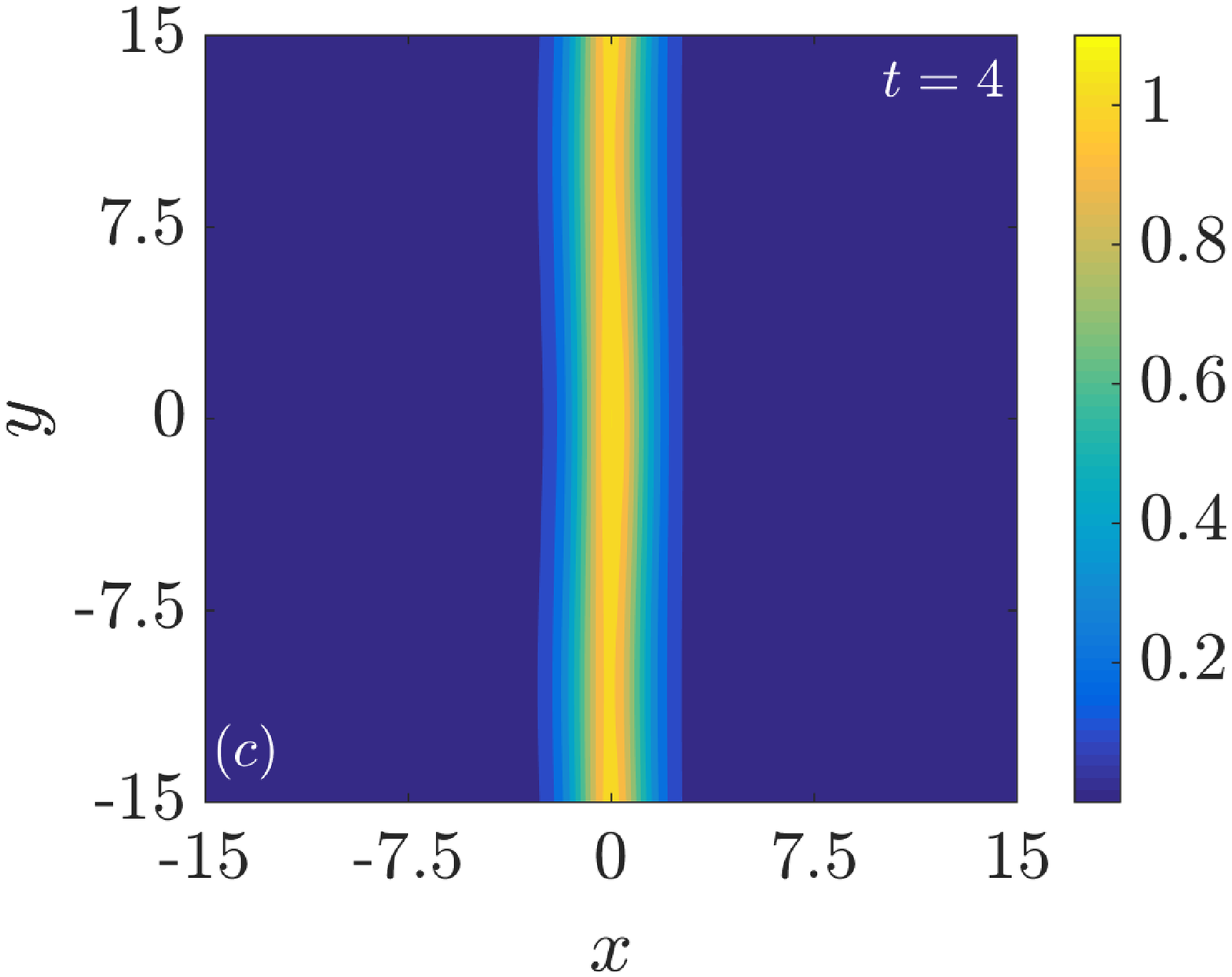}
\includegraphics[height=3.41cm]{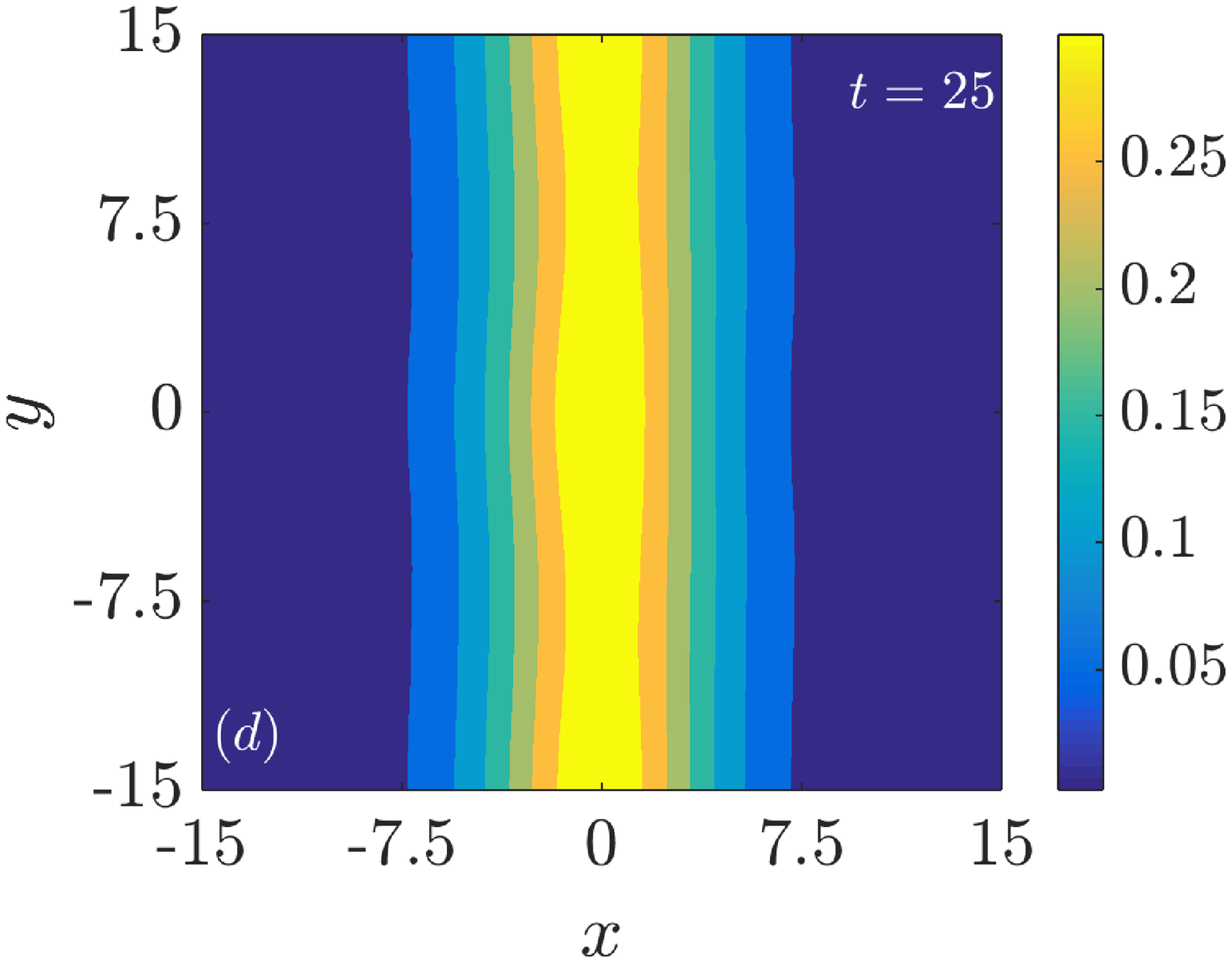}
\includegraphics[height=3.41cm]{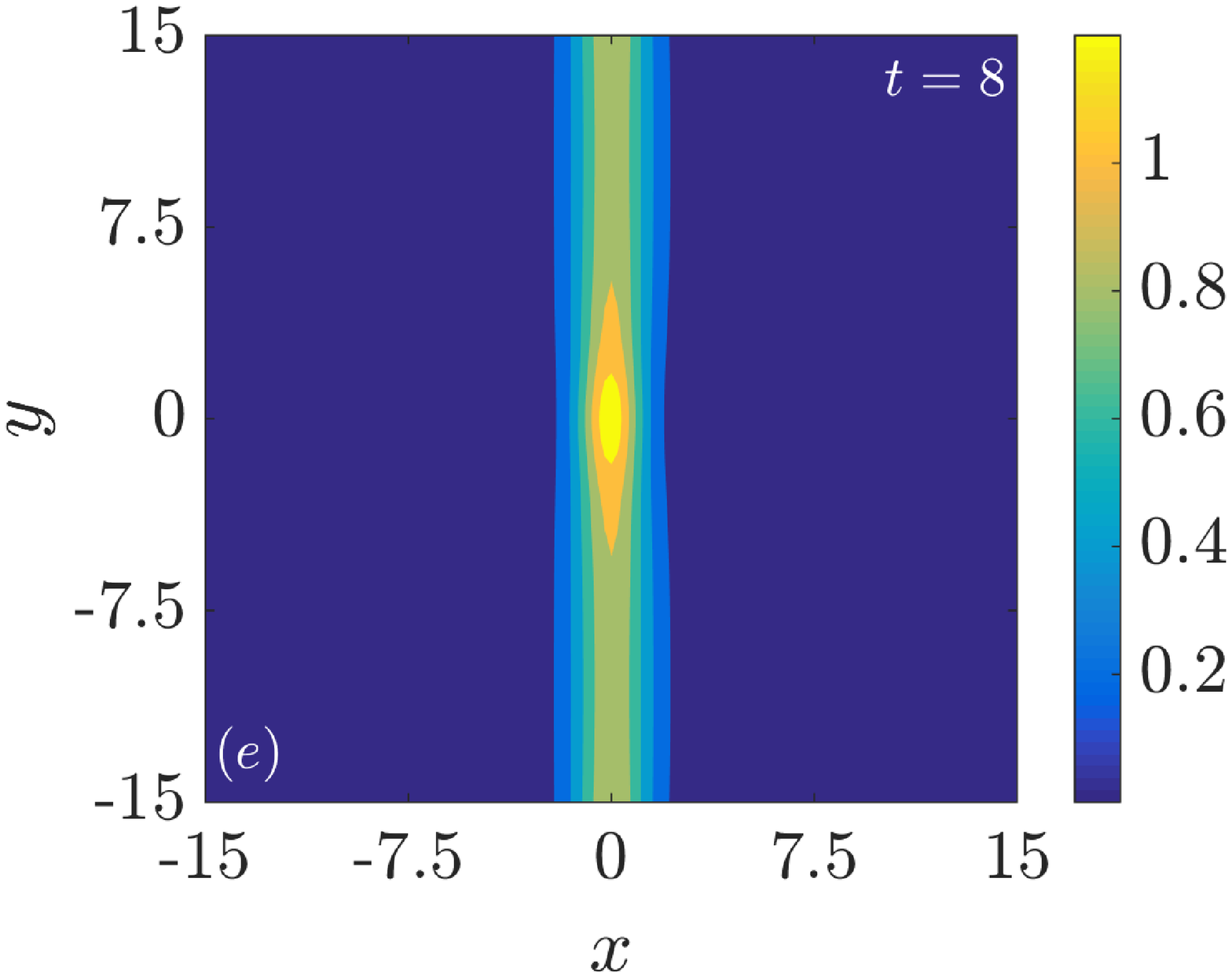}
\includegraphics[height=3.40cm]{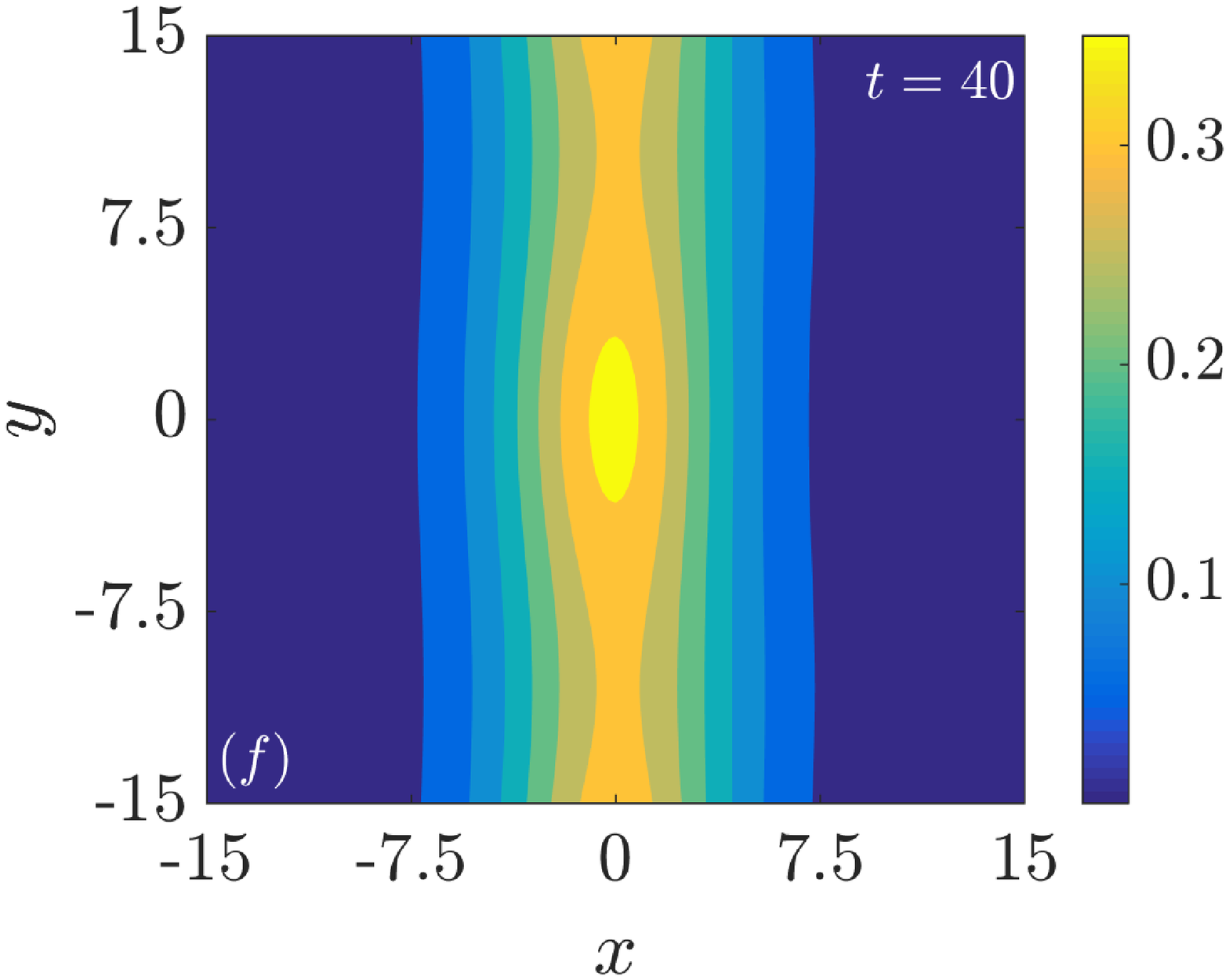}
\includegraphics[height=3.41cm]{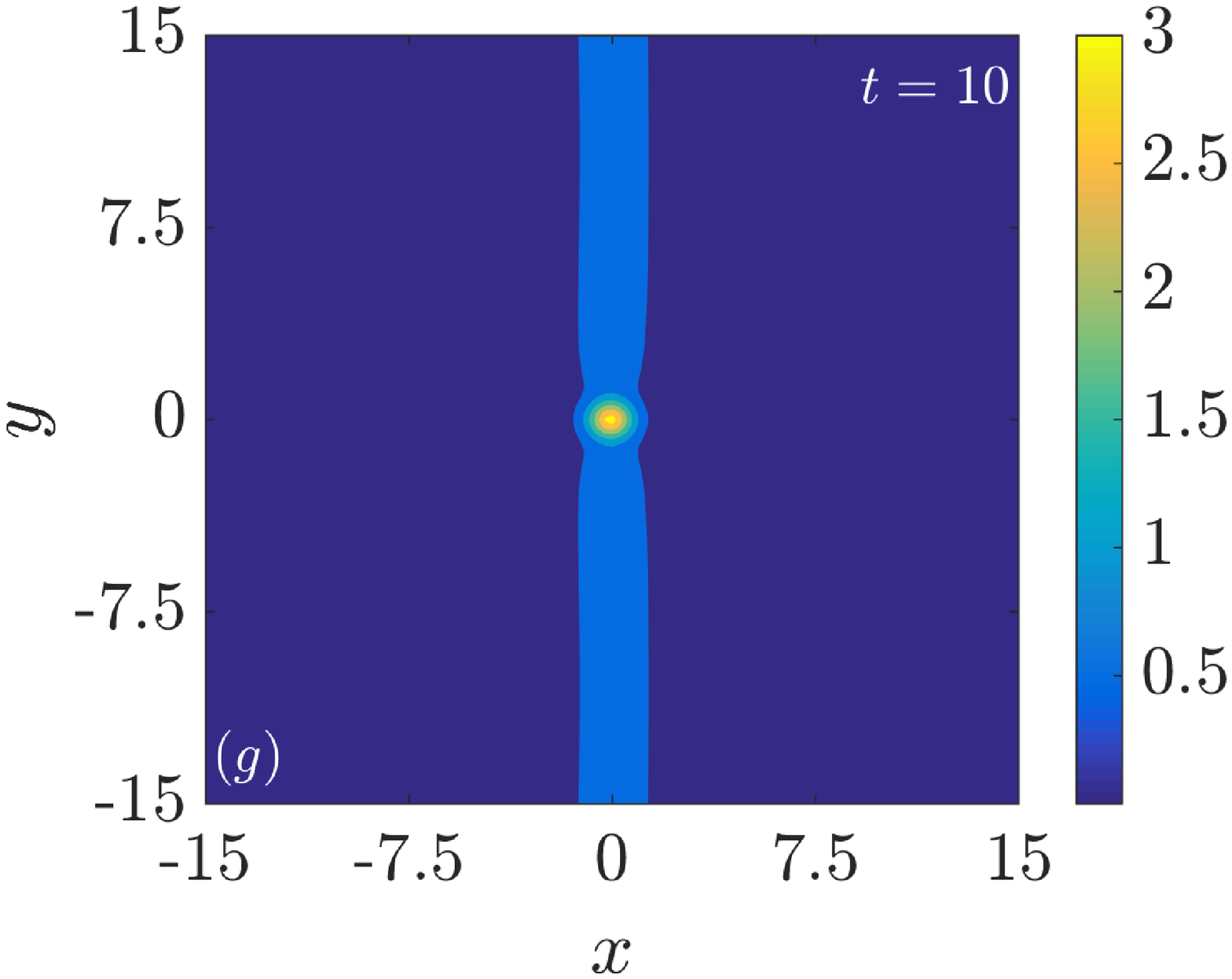}
\includegraphics[height=3.41cm]{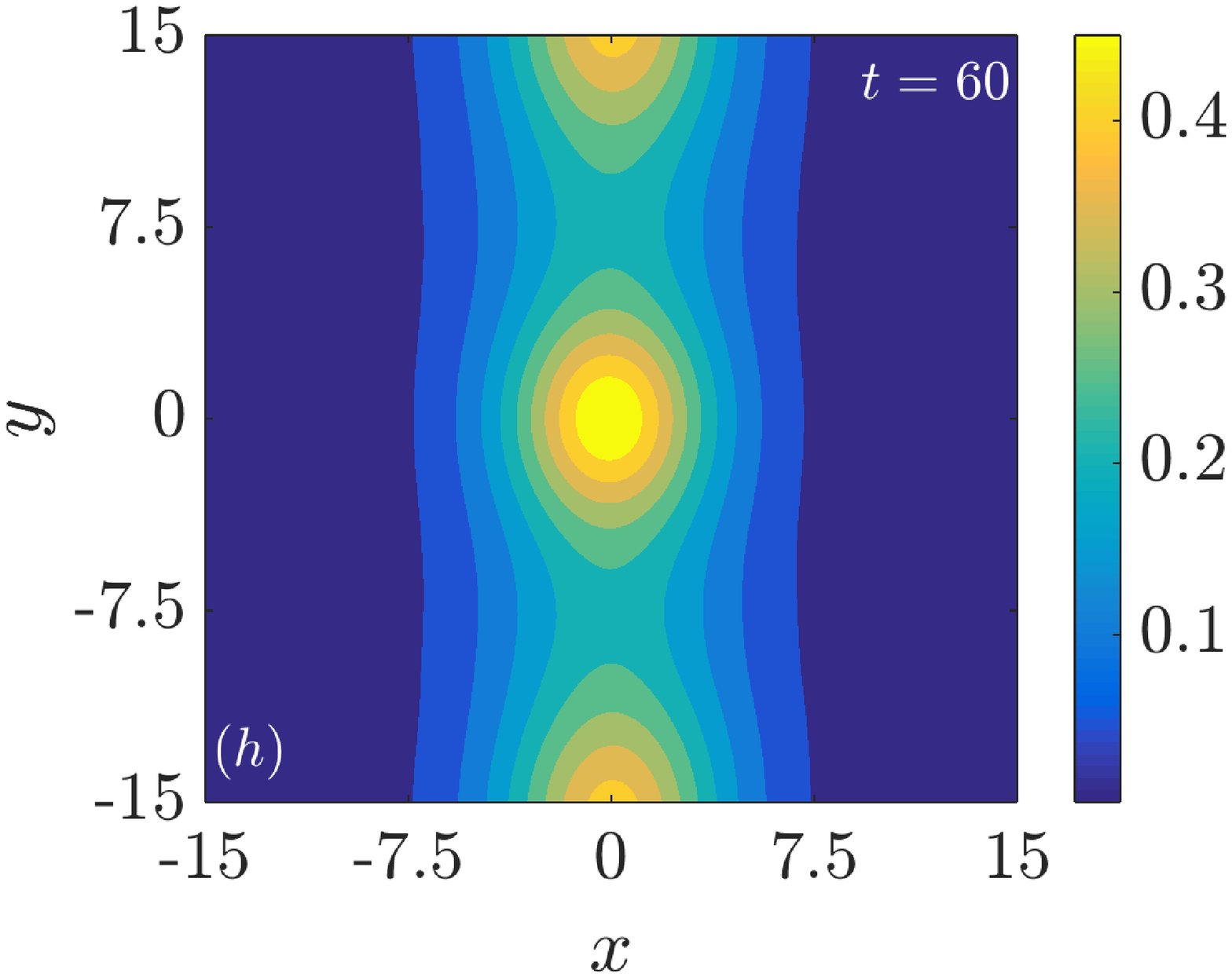}
\caption{(Color online) Contour plots showing the evolution of the soliton modulus, $|u(x,y,t)|$,
for $\nu=1$ (left panels) and $\nu=10$ (right panels); in addition, $\delta=0.1$, and other parameter values are given in Eq.~(\ref{para}). The panels (a) and (e) show
the initial condition [Eq.~(\ref{39}) for $t=0$], while the other panels show characteristic snapshots of $u$. Eventually, the soliton decays into a chain of 2D structures
(see also Fig.~\ref{dn2}).
}
\label{dn}
\end{figure}

First, we present results showcasing the instability-induced dynamics of solitons.
In Fig.~\ref{dn}, we show contour plots depicting the evolution of the soliton modulus,
$|u(x,y,t)|$, for different times. In the left panels we use the value of the nonlocality
parameter $\nu=1$, while in the right panels we showcase the larger
nonlocality
strength of $\nu=10$; other parameter values
are given in Eq.~(\ref{para}). The top panels, (a) and (b), of this figure show the initial
condition ($t=0$), as given in Eq.~(\ref{39}), while the other panels show characteristic
snapshots of $u$ for $t \ne 0$; observe that for the weaker nonlocality ($\nu=1$), the
soliton width, $1/k$ [with $k$ given in Eq.~(\ref{q0})], is shorter.

\begin{figure}[tbp!]
\centering
\includegraphics[height=3.41cm]{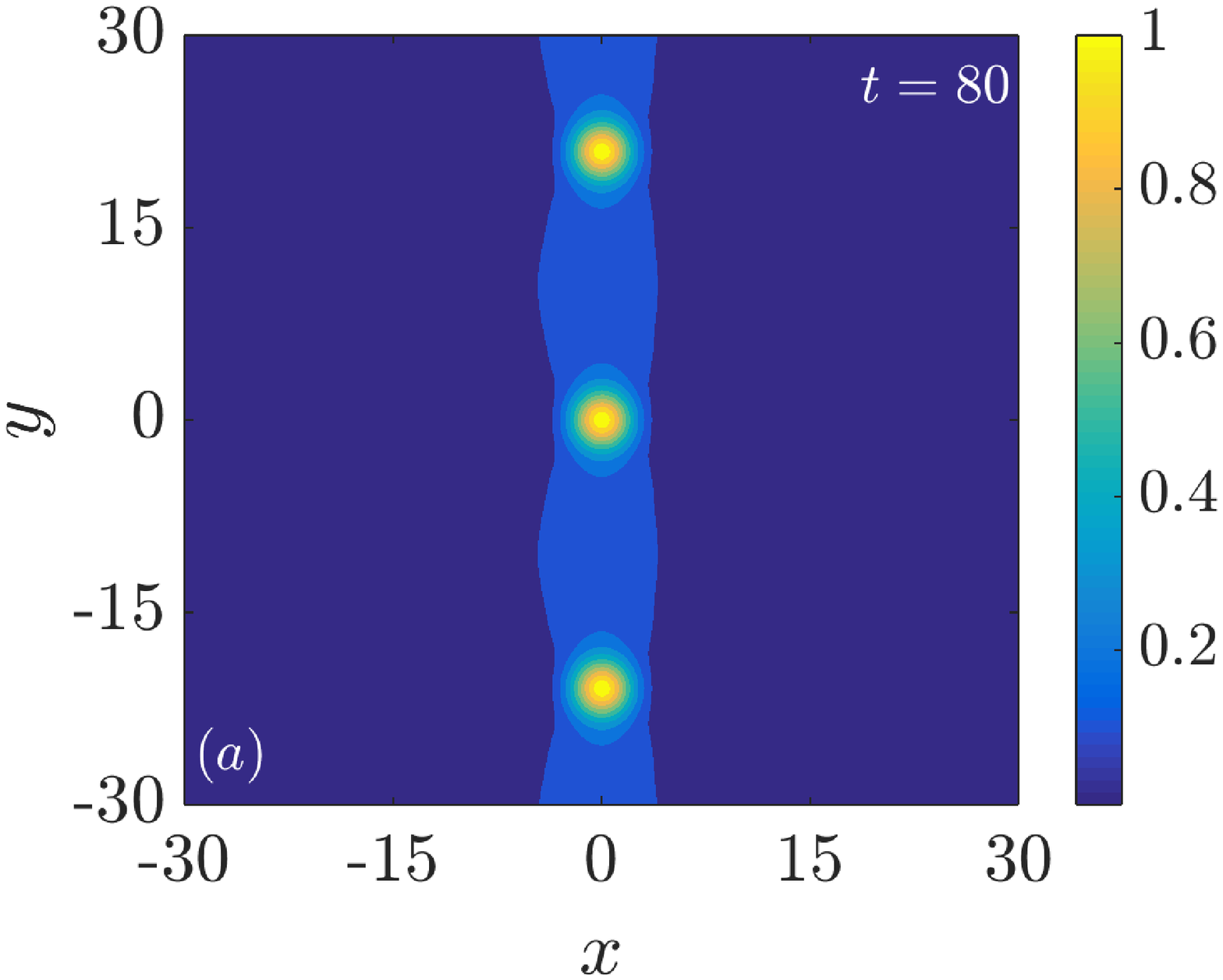}
\includegraphics[height=3.41cm]{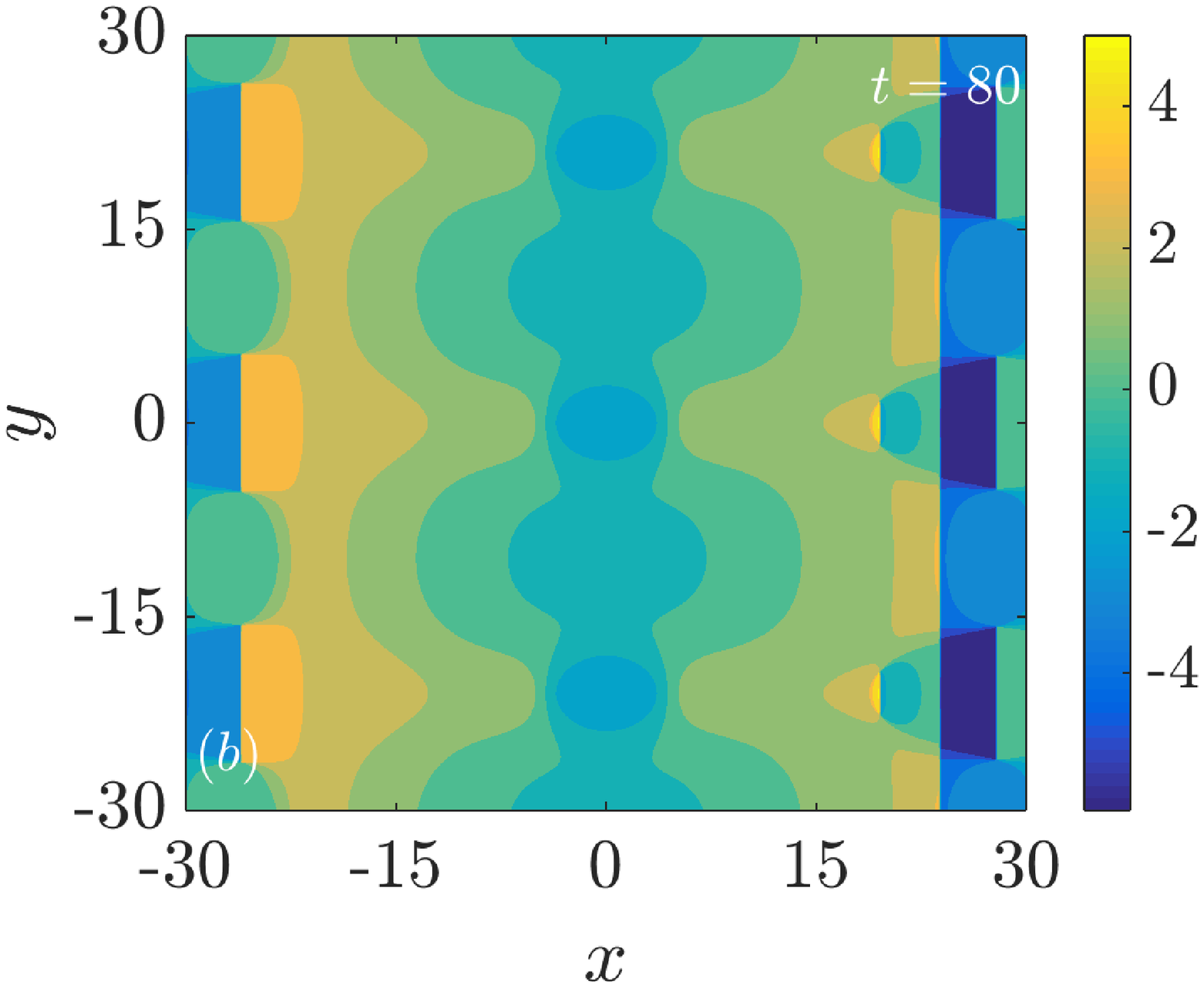}\\
\includegraphics[height=3.41cm]{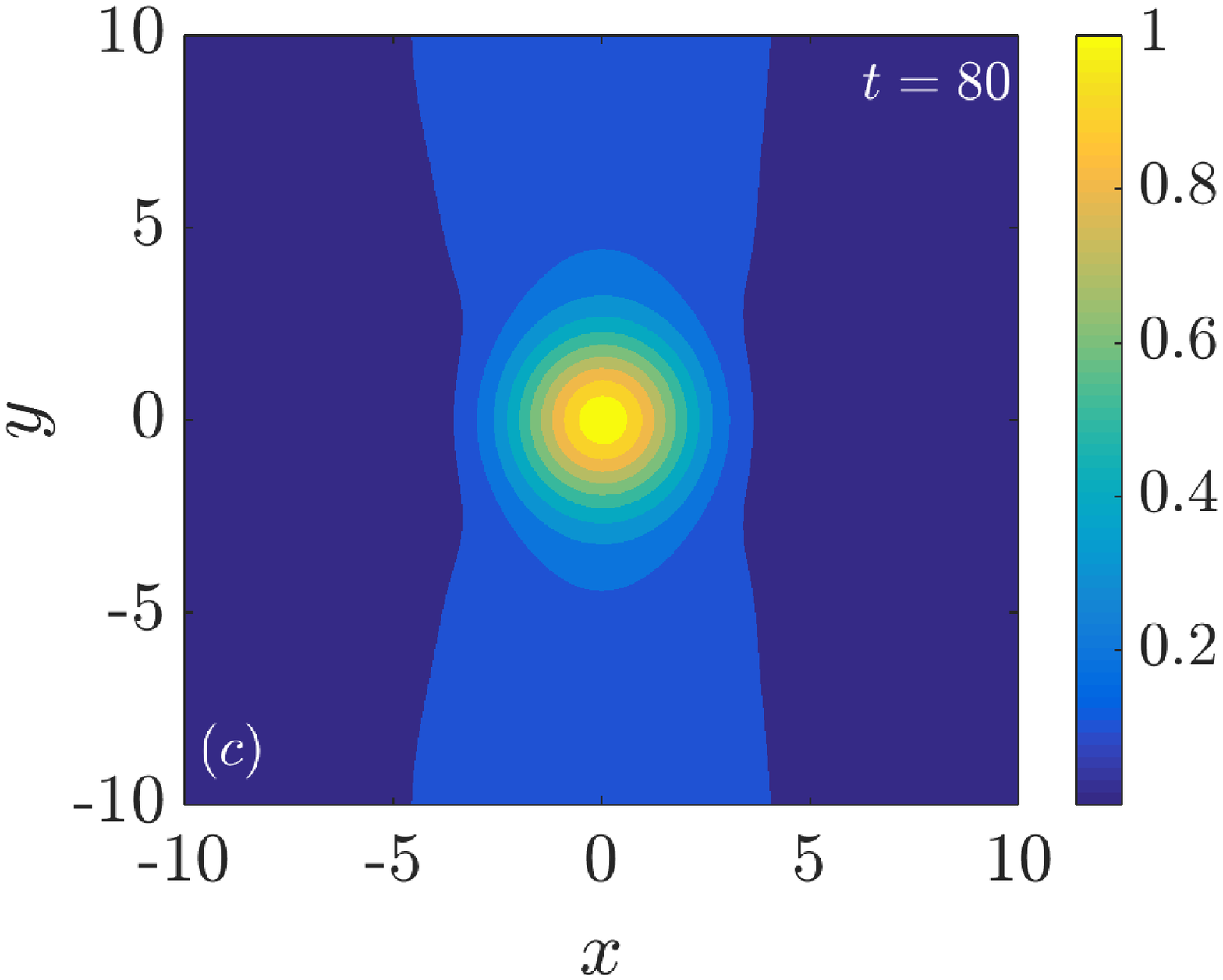}
\includegraphics[height=3.41cm]{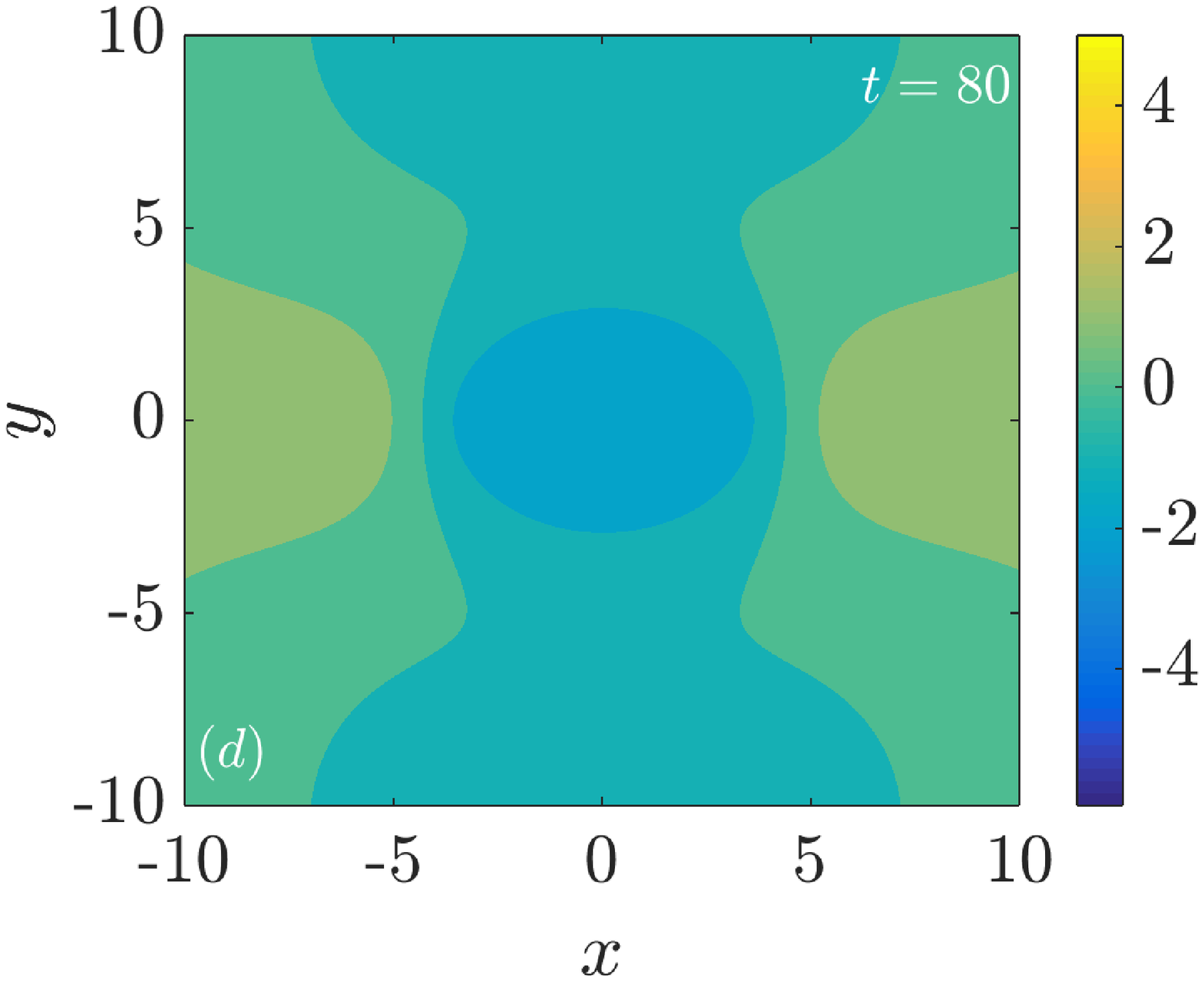}
\caption{(Color online) Contour plots showing the modulus [panel (a)] and the phase
[panel (b)] of $u(x,y,t)$ at $t=80$, for nonlocality parameter $\nu=10$; other parameters are kept fixed, as given in Eq.~(\ref{para}). This state consists of a chain of vorticity-free 2D structures, as is also seen in the bottom zoom [panels (c) and (d)].
}
\label{dn2}
\end{figure}

As is clearly seen, in both cases, the soliton stripes are prone to the instability,
which is of the necking type \cite{krz}. Nevertheless, the soliton in the setting with $\nu=10$
(right panels), takes a longer time to break up.
Hence, nonlocality leads
to a substantial suppression of the transverse instability of the soliton stripes,
similarly to what was found for the branch of solutions arising from the standard local
NLS soliton of $\nu=0$ in Ref.~\cite{lin08}. On the other hand, it is seen that eventually,
in either case, the solitons decay into a chain of 2D localized structures,
as is also shown in Fig.~\ref{dn2}; there, the modulus and the phase of such a chain
is depicted at $t=80$, for a nonlocality parameter of $\nu=10$. Notice that the phase profile of the
emerging 2D structures, depicted in the right panel of the figure, show that these waveforms are
long-lived, vorticity-free ones.

\begin{figure}[tbp!]
\centering
\includegraphics[height=7.2cm]{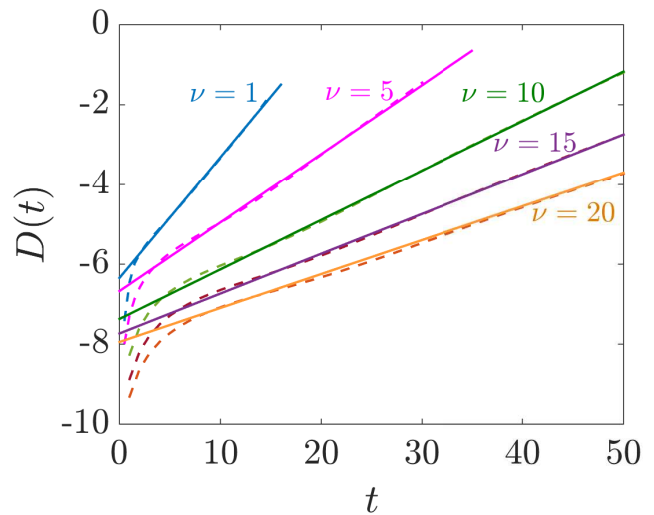}
\caption{(Color online)
The logarithm of the modulus of the difference
$D(t)=\log|u_{\rm num}(0,0,t)-u_0(0,0,t)|$, as a function of time, for different values
of the nonlocal parameter $\nu$; here, $u_{\rm num}$ is the numerical solution, and
$u_0$ is the exact analytical soliton solution. The dashed lines correspond
to the numerical results for each value of $\nu$, while the  solid lines to their
corresponding linear fits (once the instability sets in and, indeed,
after an initial transient stage). The latter are in good agreement with
the predicted growth rates 
of Eq.~(\ref{38}) (see Table~I).
}
\label{dif}
\end{figure}

At this point, it is also relevant to test the validity of the analytical estimation
for the growth rate $\Gamma=1/\tau$, with $\tau$ given by Eq.~(\ref{38}). To do this,
in Fig.~\ref{dif}, we show the logarithm of the modulus of the difference
$$D(t)=\log|u_{\rm num}(0,0,t)-u_0(0,0,t)|,$$
where $u_{\rm num}$ is the numerical solution and $u_0$ is the exact analytical
soliton solution [see Eq.~(\ref{4})], evaluated at $x=0$, $y=0$, as a function of time;
shown are curves
corresponding to different nonlocality parameters, namely $\nu=1$, $\nu=5$, $\nu=10$,
$\nu=15$, $\nu=20$. The idea here is that, subtracting the exact soliton solution
from the numerical one, one seeks to isolate the predicted exponential growth of
the soliton correction, and investigate whether it agrees with the analytical
prediction of $\exp(t/\tau)$ dependence.

The numerical results, depicted by the dashed curves, show that at the early stage
of the evolution ($t \lesssim 2$ for $\nu=1$ up to $t \lesssim 7$ for $\nu=20$),
the considered function undergoes a transient stage until the instability gets activated.
Once the latter activation materializes, the relevant plot of the logarithmic diagnostic
of choice features a linear growth. This is obviously
a signature of the exponential growth of the solution that was predicted above,
while the slopes of the pertinent straight lines should correspond to the growth rates
for the different values of $\nu$ [see Eq.~(\ref{38})]. Indeed, the slopes of the
relevant linear fits (solid lines) are close to the analytically predicted growth rates
$\Gamma = 1/\tau$ for each value of $\nu$, as shown in Table~I. As seen in the table, the
resulting relative error between the numerical result and the analytical prediction
ranges between $8\%$ to $12\%$, for all considered values of
$\nu$, signaling the good agreement between the two.

\begin{table}[h!]
\centering
\begin{tabular}{c c c c}
\hline\hline
$\nu$ &  ~~~Linear fit slope & ~~~$\Gamma=1/\tau$ & ~~~Approximate \% error \\ [0.5ex]
\hline
1  & 0.38 & 0.35 & 8 \\ 
5  & 0.17 & 0.15 & 10 \\
10 & 0.12 & 0.11 & 9 \\
15 & 0.10 & 0.09 & 10 \\
20 & 0.09 & 0.08 & 12 \\ [1ex] 
\hline
\end{tabular}
\label{table:tableI}
\caption{Comparison between the slopes of the linear fits of the numerical data of
Fig.~\ref{dif} and the analytical prediction for the growth rate, $\Gamma=1/\tau$,
for various values of $\nu$.
}
\end{table}

\begin{figure}[tbp!]
\centering
\includegraphics[height=3.417cm]{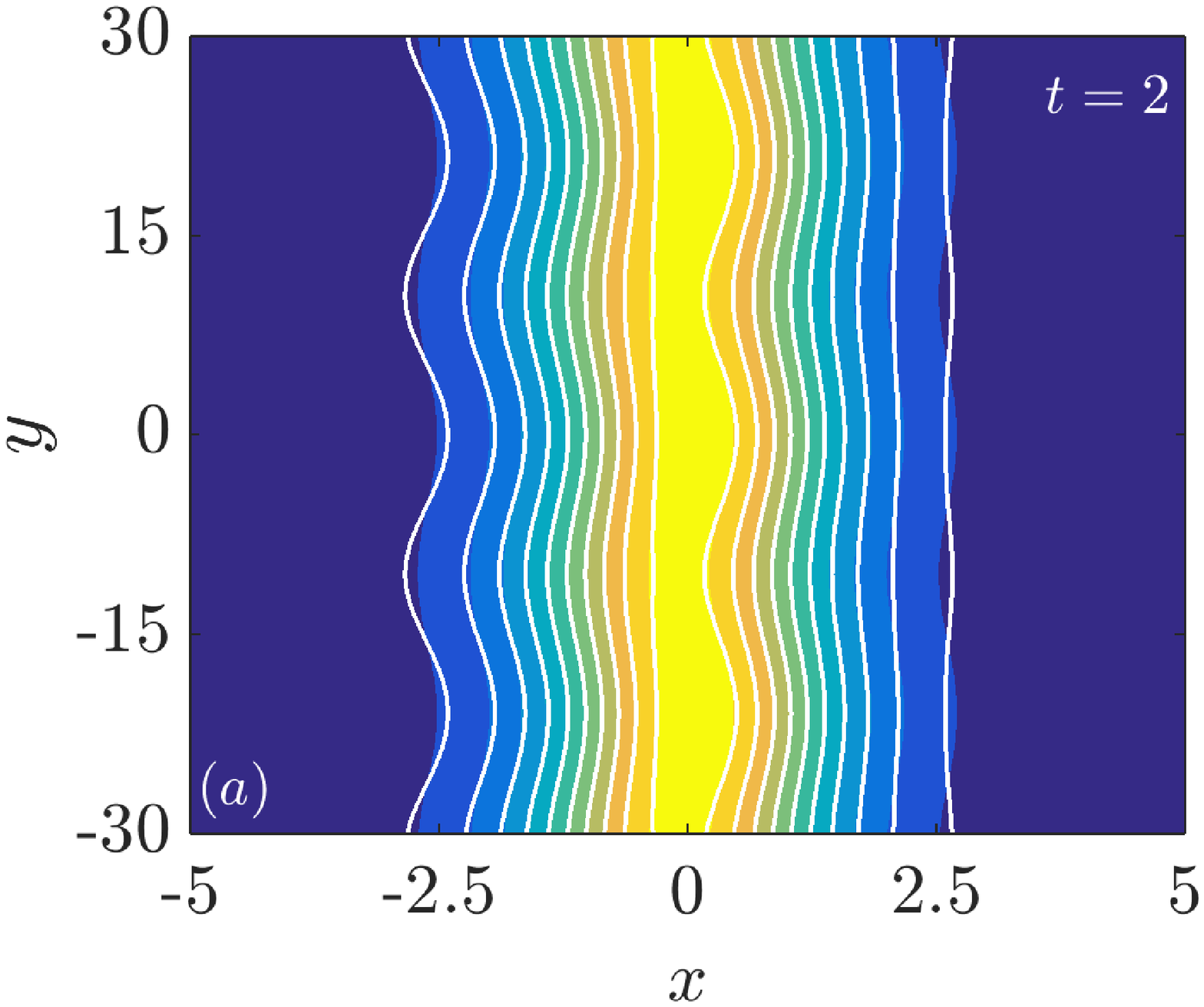}
\includegraphics[height=3.417cm]{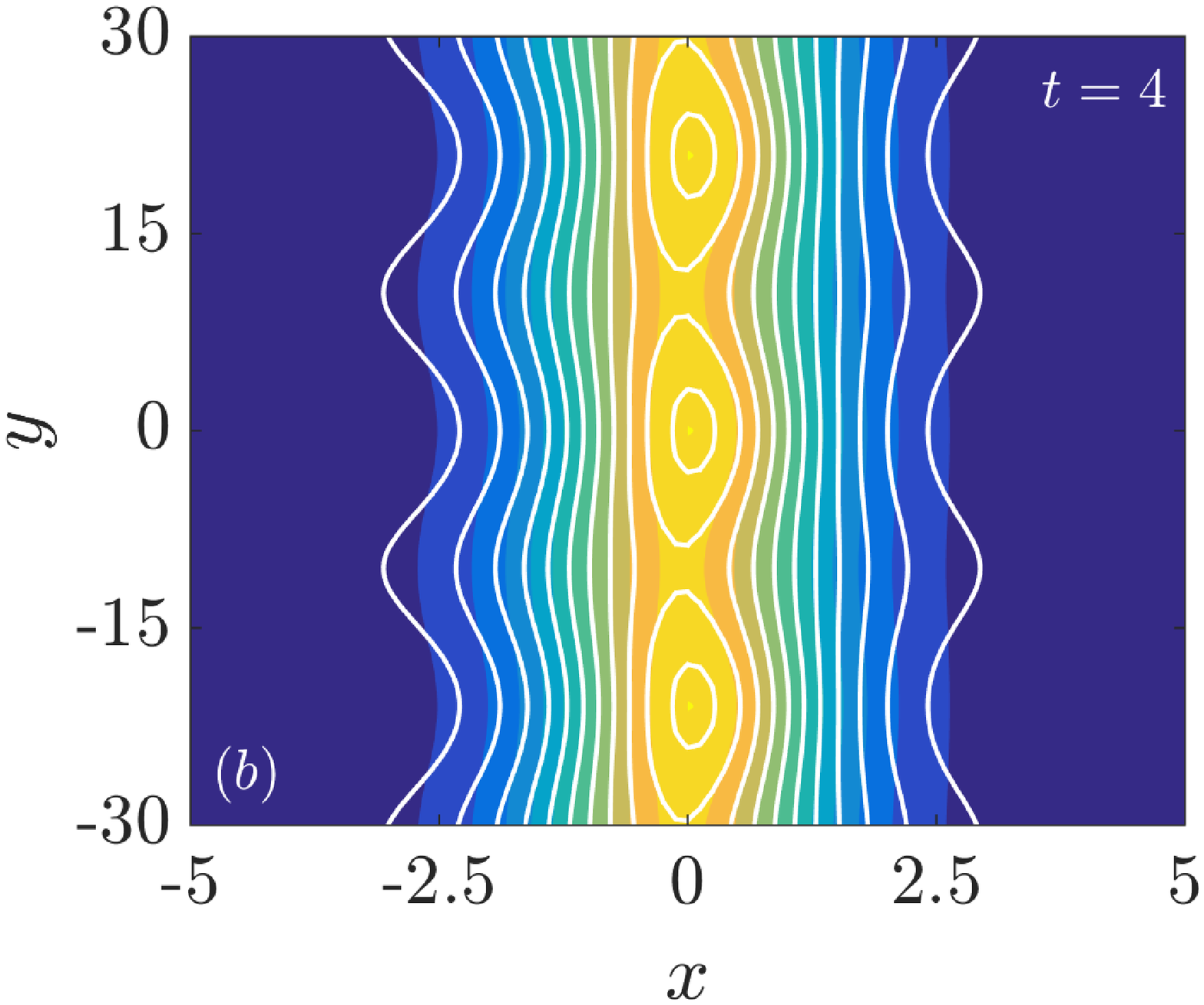}\\
\includegraphics[height=3.417cm]{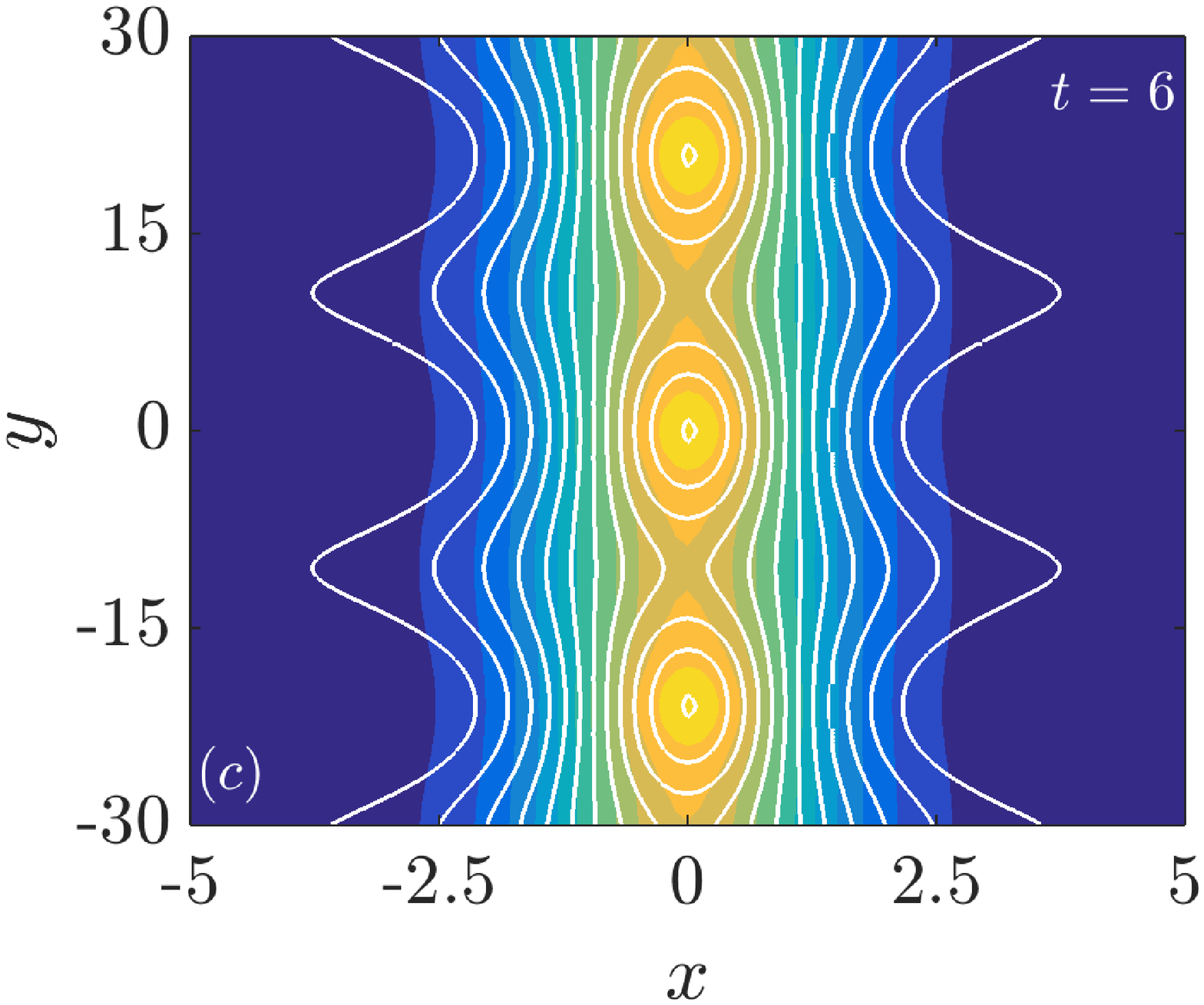}
\includegraphics[height=3.417cm]{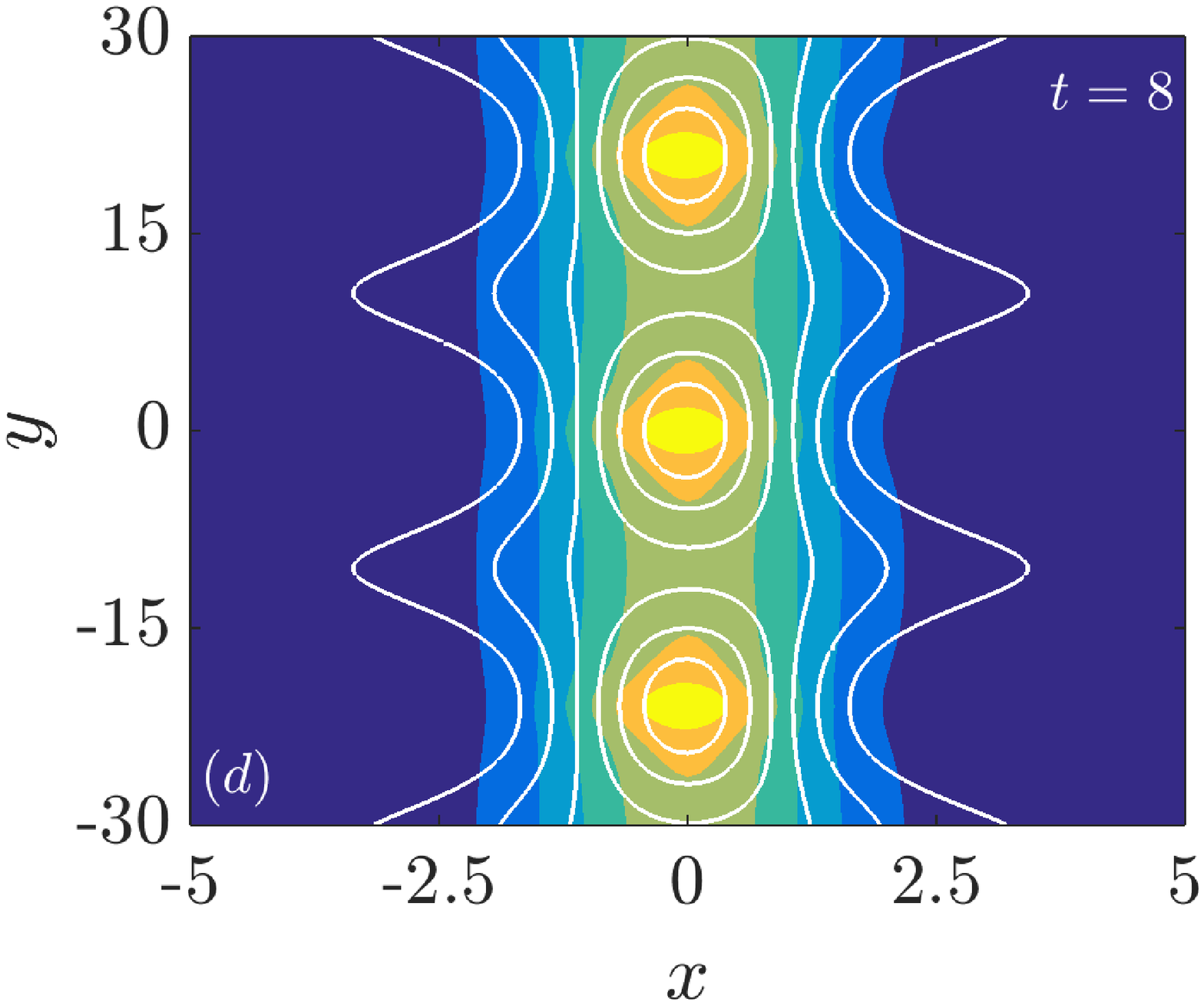}
\caption{(Color online) Solid (white) iso-contour lines of constant modulus,
corresponding to the analytical solution~(\ref{sol}), are
``superimposed'' on top of contour plots showing the modulus of the numerical
solution, for $t=2, 4, 6, 8$, and for $\nu=1$.
}
\label{compa}
\end{figure}

It is also relevant to provide an additional test for the validity of our analytical result
concerning the solution given in Eq.~(\ref{sol}), against results of the numerical simulations.
In Fig.~\ref{compa}, solid (white) iso-contour lines of constant modulus, corresponding to the
approximate analytical solution~(\ref{sol}), are ``superimposed'' on top of contour plots
showing the modulus of the numerical solution, for $\nu=1$. In all panels, (a)-(d),
corresponding to $t=2, 4, 6, 8$, a qualitative agreement between the numerical and the
analytical solution is observed, especially around the soliton
maximum. Naturally, discrepancies occur at the soliton tails as
contributions beyond our analysis of $\mathcal{O}(\epsilon)$
(such as, e.g., ones at $\mathcal{O}(\epsilon^2)$) become
progressively more important.
Notice that the discrepancy between the numerical and the analytical solution
becomes larger as time increases, due to the exponential growth of the instability,
which is only captured via the $\mathcal{O}(\epsilon)$ terms in the approximate analytical
solution of Eq.~(\ref{sol}). Finally, and as is naturally expected, past the time $t=8$
[panel~(d)], the analytical result fails, as the soliton has already been destroyed.

\section{Conclusions}
In this work, we studied the transverse dynamics and, in particular, the transverse
instability of bright soliton stripes in media with a spatially nonlocal nonlinear response.
The considered nonlocal nonlinear Schr\"odinger (NLS) model describes beam propagation
in different types of nonlocal nonlinear media, including thermal media, plasmas, and
nematic liquid crystals.

Starting with an exact 1D bright soliton solution of the system (which,
however, had no analogue in the (local) case of nonlocality parameter $\nu=0$),
we employed a direct multiscale perturbation method to study the transverse dynamics of solitons.
Assuming that the soliton's center $x_0$ and phase $\sigma_0$ become functions of
a slow time $T_1 = \epsilon t$ and a slow transverse coordinate $Y_1 = \epsilon y$
(with $0<\epsilon \ll 1$), we found the following. First, $x_0$ and $\sigma_0$ obey,
respectively, a hyperbolic and an elliptic 2nd-order PDE (with respect to $T_1$ and $Y_1$),
namely a 2nd-order wave equation and a Laplace-type equation. The solution of these
evolution equations, together with the solution for the 1st-order correction to the soliton
shape, led to an approximate solution of the original nonlocal NLS model, valid up
to $\mathcal{O}(\epsilon)$. It was found that the transverse instability,
caused by the exponential growth of the phase $\sigma_0$, is of the necking type, and
leads to the break up of solitons. The instability growth rate was found to scale with the nonlocality parameter $\nu$ according to the law $1/\sqrt{\nu}$. This
fact indicates the nonlocality-induced suppression (but not full arrest) of the
transverse instability of the bright soliton stripes, in line with results for
different solitonic structures (bearing a $\nu=0$ limit) within the model, as reported
in previous works \cite{lin08}.

Direct numerical simulations were found to be in good agreement with the analytical
predictions. As concerns the analytically found instability growth rate, it was shown
that it is in good agreement with the numerical one (past an initial transient stage),
for values of the nonlocality parameter in the interval $1 \le \nu \le 20$. In fact,
the relative percentage error between pertinent analytical and numerical results was
found to be around $10\%$ for all the cases that were considered.
In addition, the approximate analytical soliton solution
[valid up to $\mathcal{O}(\epsilon)$] was found to follow the numerical one, with the
agreement between the two being better near the the soliton center. The discrepancy
between the two, especially near the soliton tails and at later times, was attributed
to the fact that our analytical approximation cannot capture higher-order effects
[of order $\mathcal{O}(\epsilon^j)$, with $j\geq 2$], and it completely fails after the
initial solitonic stripe deforms into a sequence of two-dimensional
(non-vortical) solitonic ``blobs''.

Our work paves the way for interesting future studies. For instance, our perturbative approach
could also be applied in the case of a defocusing nonlocal nonlinearity, which supports
dark solitons, both in 1D \cite{dr1,kart1,piccardi,tph} and in 2D \cite{jphys,ol,prl}.
In such a defocusing setting, it would be interesting to study analytically the suppression
of the transverse (snaking) instability of dark soliton stripes (see relevant numerical results in
Ref.~\cite{trillo}). Furthermore, the analytical study of the transverse dynamics of
solitons in multicomponent nonlocal systems (see, e.g., Ref.~\cite{highnumber}) is another interesting and relevant theme.
This is due to the fact that that there exists a plethora of vector solitons in such settings
\cite{lee,pranem,koutso1}, while studies on the transverse dynamics of solitons are mainly numerical
ones \cite{koutso2}. It would, therefore, be particularly interesting to investigate the
combined effect of nonlocality and soliton coupling on the soliton instability dynamics.
Such studies are in progress and relevant results will be reported elsewhere.

\appendix
\section{Solution of $\mathcal{O}(\epsilon)$ perturbation equations}

Here, we provide a solution of the system of Eqs.~(\ref{14})-(\ref{16}), at the order
$\mathcal{O}(\epsilon)$ (i.e., for $j=1$). As is observed, Eq.~(\ref{15}) is decoupled
from Eqs.~(\ref{14}) and (\ref{16}) and can be solved separately. Having found the solution
of the homogeneous equation, $q_{1h}^{(\rm i)}=q_0$ [see Eq.~(\ref{26})], we seek for the
solution of the full, inhomogeneous, equation in the form:
\begin{equation}
q_{1}^{(\rm i)}(\xi,T_1,Y_1)=q_0(\xi) f(\xi,T_1,Y_1),
\label{an1}
\end{equation}
where $f(\xi)$ is an unknown function, to be determined. Substituting Eq.~(\ref{an1})
into Eq.~(\ref{15}), and employing the reduction of order method, we find: 
\begin{eqnarray}
q_{1}^{(\rm i)}=q_0 &\Bigg[ &\int \frac{1}{q_0^{2}}
\left(\int\frac{1}{k^2 d}q_0 F_{1}^{(\rm i)}d\xi\right)d\xi
\nonumber \\
&+&\int\frac{A_1(T_1,Y_1)}{q_0^2}d\xi + A_2(T_1,Y_1)\Bigg],
\label{A7}
\end{eqnarray}
where $A_1$ and $A_2$ are unknown functions of the slow variables $T_1$ and $Y_1$. Next,
imposing the boundary condition $q_{1}^{(\rm i)}\rightarrow0$ as $\xi \rightarrow\pm\infty$,
we obtain $A_1(T_i,Y_i)=0$ and we choose, without loss of generality, $A_2(T_i,Y_i)=0$ too;
indeed, the term involving $A_2$ is of the form $\epsilon A_2(T_i,Y_i)q_0$ in the
asymptotic expansion and can be absorbed in the $\mathcal{O}(1)$ solution.
This way, upon performing the relevant integrations, we derive from Eq.~(\ref{A7})
the solution~(\ref{33}).
%

The next step is to solve the system of Eqs.~(\ref{14}) and (\ref{16}). To do so, first
we solve Eq.~(\ref{16}) for the field $\theta_1$, and find:
\begin{eqnarray}
\theta_1 = \frac{1}{6g}\sqrt{\frac{d}{2\nu}}&\Bigg[&3q\sqrt{\frac{2d}{\nu}}\sigma_{0T_1}
+\left(-6g+4g\cosh^2(\xi)\right)q_1^{(\rm r)}
\nonumber \\
&-&g\cosh^2(\xi) q_{1\xi\xi}^{(\rm r)}\Bigg],
\label{eqtheta_1}
\end{eqnarray}
where we have substituted the expression $q_0$ from Eq.~(\ref{q0}).
Obviously, once $q_1^{(\rm r)}$ is found (see below), Eq.~(\ref{eqtheta_1}) can be
used for the determination of $\theta_1$.

Next, we substitute Eq.~(\ref{eqtheta_1}) into Eq.~(\ref{14}), and using the expressions
for $\theta_0$ and $q_0$ from Eqs.~(\ref{5}) and Eqs.~(\ref{q0}), we find the following
4th-order ODE for $q_{1}^{(\rm r)}$:
\begin{eqnarray}
q_{1\xi\xi\xi\xi}^{(\rm r)}&+&4\tanh(\xi)q_{1\xi\xi\xi}^{(\rm r)}-4\left(1-\sech^2(\xi)\right)q_{1\xi\xi}^{(\rm r)}
\nonumber \\
&-&16\tanh(\xi)q_{1\xi}^{(r)}-\left(16\sech^2(\xi)+72\sech^4(\xi)\right)q_{1}^{(\rm r)}
\nonumber\\
&+&\frac{12q}{g}\sqrt{\frac{2d}{\nu}}\sigma_{0T1}\sech^2(\xi)=0.
\label{eqq1r}
\end{eqnarray}
To solve the above equation, first we note that a homogeneous solution of Eq.~(\ref{eqq1r}) is
$q_{0\xi}$ [see Eq.~(\ref{26})]. Furthermore,
we can deduce that via a variation of constants method that
$q_{0\xi} \int (1/q_{0\xi}^2)d\xi$ is another homogeneous solution of
Eq.~(\ref{eqq1r}). Having at hand two homogeneous solutions,
we introduce the following transformation:
\begin{eqnarray}
q_1^{(\rm r)}(\xi)&=&\left( q_{0\xi} \int \frac{1}{q_{0\xi}^2}d\xi \right)\left( \int q_{0\xi} w(\xi) d\xi\right)
\nonumber\\
&-&q_{0\xi}\int\left(q_{0\xi}\int \frac{1}{q_{0\xi}^2} d\xi \right)w(\xi)d\xi,
\label{reduction}
\end{eqnarray}
where $w(\xi)$ is an unknown function to be determined. Substituting Eq.~(\ref{reduction})
into Eq.~(\ref{eqq1r}) we obtain the following 2nd-order ODE for $w(\xi)$:
\begin{eqnarray}
w_{\xi\xi}&+&4\tanh(\xi)w_{\xi}-8\sech^2(\xi)w
\nonumber\\
&+&\frac{12q}{g}\sqrt{\frac{2d}{\nu}}\sigma_{0T_1} \sech^2(\xi)=0.
\label{eqw}
\end{eqnarray}
It is easy to check that a partial solution of Eq.~(\ref{eqw}) is:
\begin{eqnarray}
w(\xi)=\frac{3q}{2g}\sqrt{\frac{2d}{\nu}}\sigma_{0T_1}
\label{solw}.
\end{eqnarray}
Finally, substituting Eq.~(\ref{solw}) back to Eq.~(\ref{reduction}), we derive
the solution for $q_1^{(r)}(\xi)$, namely Eq.~(\ref{34}).


\begin{thebibliography}{10}

\bibitem{Infeld}
E.~Infeld and G.~Rowlands.
\newblock {\em Nonlinear Waves, Solitons and Chaos}.
\newblock Cambridge University Press, 1990.

\bibitem{zo}
V.~E. Zakharov and L.~A. Ostrovsky.
\newblock Modulation instability: The beginning.
\newblock {\em Physica D}, 238:540--548, 2009.

\bibitem{ZR}
V.~E. Zakharov and A.~M. Rubenchik.
\newblock Instability of waveguides and solitons in nonlinear media.
\newblock {\em Sov. Phys. JETP}, 38:494–500, 1974.

\bibitem{han}
S.~J. Han.
\newblock Stability of envelope waves.
\newblock {\em Phys. Rev. A}, 20:2568–2573, 1979.

\bibitem{kuz}
E.~A. Kuznetsov and S.~K. Turitsyn.
\newblock Instability and collapse of solitons in media with a defocusing
  nonlinearity.
\newblock {\em J. Exp. Theor. Phys.}, 67:1583–1588, 1988.

\bibitem{peli1}
D.~E. Pelinovsky, Yu.~A. Stepanyants, and Yu.~S. Kivshar.
\newblock Self-focusing of plane dark solitons in nonlinear defocusing media.
\newblock {\em Phys. Rev. E}, 51:5016--5026, 1995.

\bibitem{krz}
E.~A. Kuznetsov, A.~M. Rubenchik, and V.~E. Zakharov.
\newblock Soliton stability in plasmas and hydrodynamics.
\newblock {\em Phys. Rep.}, 142:103–165, 1986.

\bibitem{pelirev}
Yu.~S. Kivshar and D.~E. Pelinovsky.
\newblock Self-focusing and transverse instabilities of solitary waves.
\newblock {\em Phys. Rep.}, 331:117--195, 2000.

\bibitem{yang}
J.~Yang.
\newblock {\em Nonlinear Waves in Integrable and Nonintegrable Systems}.
\newblock SIAM, 2010.

\bibitem{anderson}
B.~P. Anderson, P.~C. Haljan, C.~A. Regal, D.~L. Feder, L.~A. Collins, C.~W.
  Clark, and E.~A. Cornell.
\newblock Watching dark solitons decay into vortex rings in a bose-einstein
  condensate.
\newblock {\em Phys. Rev. Lett.}, 86:2926--2929, Apr 2001.

\bibitem{bernard}
S.-P. Gorza, B.~Deconinck, Ph. Emplit, T.~Trogdon, and M.~Haelterman.
\newblock Experimental demonstration of the oscillatory snake instability of
  the bright soliton of the $(2+1)\mathbf{D}$ hyperbolic nonlinear
  schr\"odinger equation.
\newblock {\em Phys. Rev. Lett.}, 106:094101, Mar 2011.

\bibitem{bernard2}
Simon-Pierre Gorza, Bernard Deconinck, Thomas Trogdon, Philippe Emplit, and
  Marc Haelterman.
\newblock Neck instability of bright solitons in normally dispersive kerr
  media.
\newblock {\em Opt. Lett.}, 37(22):4657--4659, Nov 2012.

\bibitem{muss1}
Z.~H. Musslimani, M.~Segev, A.~Nepomnyashchy, and Yu.~S. Kivshar.
\newblock Suppression of transverse instabilities for vector solitons.
\newblock {\em Phys. Rev. E}, 60:R1170–R1173, 1999.

\bibitem{muss2}
Z.~H. Musslimani and J.~Yang.
\newblock Transverse instability of strongly coupled dark-bright manakov vector
  solitons.
\newblock {\em Opt. Lett.}, 26:1981--1983, 2001.

\bibitem{ana}
C.~Anastassiou, M.~Soljacic, M.~Segev, E.~D. Eugenieva, D.~N. Christodoulides,
  D.~Kip, Z.~H. Musslimani, and J.~P. Torres.
\newblock Eliminating the transverse instabilities of kerr solitons.
\newblock {\em Phys. Rev. Lett.}, 85:4888–4891, 2000.

\bibitem{y1}
J.~Yang.
\newblock Transversely stable soliton trains in photonic lattice.
\newblock {\em Phys. Rev. A}, 84:033840, 2011.

\bibitem{y2}
J.~Yang, D.~Gallardo, A.~Miller, and Z.~Chen.
\newblock Elimination of transverse instability in stripe solitons by
  one-dimensional lattices.
\newblock {\em Opt. Lett.}, 37:1571–1573, 2012.

\bibitem{Ma}
Manjun Ma, R.~Carretero-Gonz{\'a}lez, P.~G. Kevrekidis, D.~J. Frantzeskakis,
  and B.~A. Malomed.
\newblock Controlling the transverse instability of dark solitons and
  nucleation of vortices by a potential barrier.
\newblock {\em Phys. Rev. A}, 82:023621, 2010.

\bibitem{litvak}
A.~G. Litvak, V.~A. Mironov, G.~M. Fraiman, and A.~D. Yunakovskii.
\newblock Thermal self-effect of wave beams in a plasma with a nonlocal
  nonlinearity.
\newblock {\em Sov. J. Plasma Phys.}, 1:60--71, 1975.

\bibitem{vap}
D.~Suter and T.~Blasberg.
\newblock Stabilization of transverse solitary waves by a nonlocal response of
  the nonlinear medium.
\newblock {\em Phys. Rev. A}, 48:4583--4587, 1993.

\bibitem{rot}
C.~Rotschild, T.~Carmon, O.~Cohen, O.~Manela, and M.~Segev.
\newblock Solitons in nonlinear media with an infinite range of nonlocality:
  first observation of coherent elliptic solitons and of vortex-ring solitons.
\newblock {\em Phys. Rev. Lett.}, 95:213904, 2005.

\bibitem{ass0}
C.~Conti, M.~Peccianti, and G.~Assanto.
\newblock Route to nonlocality and observation of accessible solitons.
\newblock {\em Phys. Rev. Lett.}, 91:073901, 2003.

\bibitem{santos}
P.~Pedri and L.~Santos.
\newblock Two-dimensional bright solitons in dipolar {Bose-Einstein}
  condensates.
\newblock {\em Phys. Rev. Lett.}, 95:200404, 2005.

\bibitem{lin08}
YuanYao Lin, Ray-Kuang Lee, and Yu.~S. Kivshar.
\newblock Suppression of soliton transverse instabilities in nonlocal nonlinear
  media.
\newblock {\em J. Opt. Soc. Am. B}, 2008:576--581, 2008.

\bibitem{tursolo}
S.~K. Turitsyn.
\newblock Spatial dispersion of nonlinearity and stability of many dimensional
  solitons.
\newblock {\em Theor. Math. Phys.}, 64:797--801, 1985.

\bibitem{krol1}
W.~Krolikowski, O.~Bang, N.~I. Nikolov, D.~Neshev, J.~Wyller, J.~J. Rasmussen,
  and D.~Edmundson.
\newblock Modulational instability, solitons and beam propagation in spatially
  nonlocal nonlinear media.
\newblock {\em J. Opt. B: Quantum Semiclass. Opt.}, 6:S288--S294, 2004.

\bibitem{Mih2}
D.~Mihalache, D.~Mazilu, F.~Lederer, B.~A. Malomed, Y.~V. Kartashov, L.-C.
  Crasovan, and L.~Torner.
\newblock Three-dimensional spatiotemporal optical solitons in nonlocal
  nonlinear media.
\newblock {\em Phys. Rev. E}, 73:025601(R), 2006.

\bibitem{Mih}
D.~Mihalache.
\newblock Multidimensional solitons and vortices in nonlocal nonlinear optical
  media.
\newblock {\em Rom. Rep. Phys.}, 59:515--522, 2007.

\bibitem{dr1}
A.~Dreischuh, D.~N. Neshev, D.~E. Petersen, O.~Bang, and W.~Krolikowski.
\newblock Observation of attraction between dark solitons.
\newblock {\em Phys. Rev. Lett.}, 96:043901, 2006.

\bibitem{kart1}
Y.~V. Kartashov and L.~Torner.
\newblock Gray spatial solitons in nonlocal nonlinear media.
\newblock {\em Opt. Lett.}, 32:946--948, 2007.

\bibitem{piccardi}
A.~Piccardi, A.~Alberucci, N.~Tabiryan, and G.~Assanto.
\newblock Dark nematicons.
\newblock {\em Opt. Lett.}, 36:1356--1358, 2011.

\bibitem{tph}
T.~P. Horikis.
\newblock Small-amplitude defocusing nematicons.
\newblock {\em J. Phys. A: Math. Theor.}, 48:02FT01, 2015.

\bibitem{trillo}
A.~Armaroli and S.~Trillo.
\newblock Suppression of transverse instabilities of dark solitons and their
  dispersive shock waves.
\newblock {\em Phys. Rev. A}, 80:053803, 2009.

\bibitem{liq1}
N.~Ghofraniha, C.~Conti, G.~Ruocco, and S.~Trillo.
\newblock Shocks in nonlocal media.
\newblock {\em Phys. Rev. Lett.}, 99:043903, 2007.

\bibitem{liq2}
C.~Conti, A.~Fratalocchi, M.~Peccianti, G.~Ruocco, and S.~Trillo.
\newblock Observation of a gradient catastrophe generating solitons.
\newblock {\em Phys. Rev. Lett.}, 102:083902, 2009.

\bibitem{pl1}
A.~G. Litvak, V.~A. Mironov, G.~M. Fraiman, and A.~D. Yunakovskii.
\newblock Thermal self-effect of wave beams in a plasma with a nonlocal
  nonlinearity.
\newblock {\em Sov. J. Plasma Phys.}, 1:60--71, 1975.

\bibitem{pl2}
A.~I. Yakimenko, Y.~A. Zaliznyak, and Yu.~S. Kivshar.
\newblock Stable vortex solitons in nonlocal self-focusing nonlinear media.
\newblock {\em Phys. Rev. E}, 71:065603(R), 2005.

\bibitem{pec}
M.~Peccianti and G.~Assanto.
\newblock Nematicons.
\newblock {\em Phys. Rep.}, 516:147--208, 2012.

\bibitem{alb}
A.~Alberucci and G.~Assanto.
\newblock Modeling nematicon propagation.
\newblock {\em Mol. Cryst. Liq. Cryst.}, 572:2--12, 2013.

\bibitem{pes}
M.~Z. Pesenson.
\newblock Nonlinear waves traveling upon a front of solitons.
\newblock {\em Phys. Fluids A}, 3:3001--3006, 1991.

\bibitem{MJA.segur}
M.~J. Ablowitz and H.~Segur.
\newblock {\em Solitons and the Inverse Scattering Transform}.
\newblock SIAM, 1981.

\bibitem{ass1}
G.~Assanto.
\newblock {\em Nematicons: {S}patial Optical Solitons in Nematic Liquid
  Crystals}.
\newblock New Jersey: Wiley-Blackwell, 2012.

\bibitem{pana2}
J.~M.~L. MacNeil, N.~F. Smyth, and G.~Assanto.
\newblock Exact and approximate solutions for solitary waves in nematic liquid
  crystals.
\newblock {\em Physica D}, 284:1--15, 2014.

\bibitem{kath}
W.~L. Kath and N.~F. Smyth.
\newblock Soliton evolution and radiation loss for the nonlinear
  {Schr\"odinger} equation.
\newblock {\em Phys. Rev. E}, 51:1484--1492, 1995.

\bibitem{mjads}
M.~J. Ablowitz, S.~D. Nixon, T.~P. Horikis, and D.~J. Frantzeskakis.
\newblock Perturbations of dark solitons.
\newblock {\em Proc. R. Soc. A}, 467:2597–2621, 2011.

\bibitem{kassam}
A.~Kassam and L.~N. Trefethen.
\newblock Fourth-order time stepping for stiff {PDE}s.
\newblock {\em SIAM J. Sci. Comput.}, 26:1214--1233, 2005.

\bibitem{jphys}
T.~P. Horikis and D.~J. Frantzeskakis.
\newblock Asymptotic reductions and solitons of nonlocal nonlinear
  {Schr\"odinger} equations.
\newblock {\em J. Phys. A: Math. Theor.}, 49:205202, 2016.

\bibitem{ol}
T.~P. Horikis and D.~J. Frantzeskakis.
\newblock Ring dark and antidark solitons in nonlocal media.
\newblock {\em Opt. Lett.}, 41:583--586, 2016.

\bibitem{prl}
T.~P. Horikis and D.~J. Frantzeskakis.
\newblock Light meets water in nonlocal media: {Surface} tension in optics.
\newblock {\em Phys. Rev. Lett.}, 118:243903, 2017.

\bibitem{highnumber}
A.~Alberucci, M.~Peccianti, G.~Assanto, A.~Dyadyusha, and M.~Kaczmarek.
\newblock Two-color vector solitons in nonlocal media.
\newblock {\em Phys. Rev. Lett.}, 97:153903, 2006.

\bibitem{lee}
Y.~Lin and R.-K. Lee.
\newblock Dark-bright soliton pairs in nonlocal nonlinear media.
\newblock {\em Opt. Express}, 15:8781, 2007.

\bibitem{pranem}
T.~P. Horikis and D.~J. Frantzeskakis.
\newblock Vector nematicons: Coupled spatial solitons in nematic liquid
  crystals.
\newblock {\em Phys. Rev. A}, 94:053805, 2016.

\bibitem{koutso1}
G.~N. Koutsokostas, T.~P. Horikis, D.~J. Frantzeskakis, B.~Prinari, and
  G.~Biondini.
\newblock Multiscale expansions and vector solitons of a two-dimensional
  nonlocal nonlinear {Schr\"odinger} system.
\newblock {\em Stud. Appl. Math.}, 145:739--764, 2020.

\bibitem{koutso2}
G.~N. Koutsokostas, T.~P. Horikis, D.~J. Frantzeskakis, B.~Prinari, and
  G.~Biondini.
\newblock Transverse dynamics of vector solitons in defocusing nonlocal media.
\newblock {\em Eur. Phys. J. Plus}, 135:546, 2020.

\end{thebibliography}

\end{document}